\documentclass[showpacs,amsmath,amssymb,aps,twocolumn,superscriptaddress,prx]{revtex4-2} 
\usepackage{mathtools}
\usepackage{algorithm}
\usepackage{algpseudocode}
\usepackage[pdftex]{graphicx} \graphicspath{{}}% Include figure files
\usepackage{grffile} % change algorithm for searching thru graphics extensions

\usepackage{comment}

\usepackage{float}
\usepackage{dcolumn}% Align table columns on decimal point
\usepackage{bm}% bold math
\usepackage[pdftex,colorlinks=true]{hyperref}
\hypersetup{
	colorlinks=true,
	linkcolor=blue,
	urlcolor=cyan,
}

\usepackage{color}
\definecolor{ForestGreen}{RGB}{34, 139, 34}

\newcommand{\ket}[1]{\left | #1 \right\rangle}

\begin{document}
	
	\title{Making Trotterization adaptive and energy-self-correcting for NISQ devices and beyond}
	
	\author{Hongzheng Zhao}
	\email{hzhao@pks.mpg.de}
	\affiliation{Max Planck Institute for the Physics of Complex Systems, N\"{o}thnitzer Str.~38, 01187 Dresden, Germany}
	
	\author{Marin Bukov}
	\affiliation{Max Planck Institute for the Physics of Complex Systems, N\"{o}thnitzer Str.~38, 01187 Dresden, Germany}
	\affiliation{Department of Physics, St.~Kliment Ohridski University of Sofia, 5 James Bourchier Blvd, 1164 Sofia, Bulgaria}
	
	\author{Markus Heyl}
	\affiliation{Max Planck Institute for the Physics of Complex Systems, N\"{o}thnitzer Str.~38, 01187 Dresden, Germany}
	\affiliation{Theoretical Physics III, Center for Electronic Correlations and Magnetism,
		Institute of Physics, University of Augsburg, 86135 Augsburg, Germany}
	
	\author{Roderich Moessner}
	\affiliation{Max Planck Institute for the Physics of Complex Systems, N\"{o}thnitzer Str.~38, 01187 Dresden, Germany}

	\date{\today}
	
	\begin{abstract}
		Simulation of continuous time evolution requires time discretization on both classical and quantum computers.
		A finer time step improves simulation precision, but it
		inevitably leads to increased computational efforts. This
		is particularly costly for today's noisy intermediate scale
		quantum computers, where notable gate imperfections
		limit the circuit depth that can be executed at a given
		accuracy. Classical adaptive solvers are well-developed to
		save numerical computation times. However, it remains
		an outstanding challenge to make optimal usage of the
		available quantum resources by means of adaptive time steps.
		Here, we introduce a quantum algorithm to solve this
		problem, providing a controlled solution of the quantum
		many-body dynamics of local observables. The key conceptual element of our algorithm is a feedback loop which
		self-corrects the simulation errors by adapting time steps,
		thereby significantly outperforming conventional Trotter
		schemes on a fundamental level and reducing the circuit
		depth. It even allows for a controlled asymptotic long-time
		error, where usual Trotterized dynamics is facing difficulties.
		Another key advantage of our quantum algorithm is that any desired conservation law can be included in the self-correcting feedback loop, which has potentially a wide range of applicability. We demonstrate the capabilities by enforcing gauge invariance which is crucial for
		a faithful and long-sought quantum simulation of lattice gauge theories. Our algorithm can
		be potentially useful on a more general level whenever time discretization is involved concerning, for instance, also numerical approaches
		based on time-evolving block decimation methods.
	\end{abstract}
	\maketitle

	\let\oldaddcontentsline\addcontentsline % Store \addcontentsline
	\renewcommand{\addcontentsline}[3]{} % Make \addcontentsline a no-op 
	
\section{Introduction}
Quantum computers hold the promise to outperform their classical counterparts in certain computational tasks. Among others, emulating quantum many-body systems in and out of equilibrium attract considerable attention~\cite{georgescu2014quantum,preskill2018quantum}. 
State-of-the-art quantum devices for digital quantum simulation (DQS) such as trapped ions~\cite{blatt2012quantum,monroe2021programmable,dumitrescu2022dynamical}, superconducting circuits~\cite{salathe2015digital,satzinger2021realizing,dborin2022simulating} and Rydberg platforms~\cite{jaksch2000fast,saffman2010quantum,levine2018high}, have recently realized condensed matter models~\cite{garttner2017measuring,smith2019simulating,landsman2019verified,sun2021perturbative,barratt2021parallel,han2021experimental,de2021quantum,pastori2022characterization,kamakari2022digital,mi2022time,green2022experimental}, simulation of
molecular energies~\cite{o2016scalable,mcardle2020quantum} and lattice gauge theories (LGTs) with the potential to answer longstanding questions in high energy physics~\cite{schweizer2019floquet,lamm2019general,yang2020observation,vovrosh2021confinement,tan2021domain,mildenberger2022probing,nguyen2022digital,klco2022standard}.

In this spirit, we consider a general quantum many-body system described by a Hamiltonian $H{=}\sum_{j=1}^N H_j$ with $N$ terms, whose dynamics we would like to simulate. In DQS, we assert that the time evolution generated by each individual term $H_j$ can be implemented on the simulator, but not the total Hamiltonian $H$.
The difficulty in DQS arises from the noncommutativity $[H_i,H_j]\neq0$ ($i\neq j$), which is a defining feature of quantum mechanics.
The basic idea behind DQS is to decompose the target time evolution operator $U=\exp(-itH)$ into a series of elementary few-body quantum gates. Over a short time step $\delta t$, we can approximate $U$ by the digitized
$U_T(\delta t) = \prod_{j=1}^N \exp(-i\delta tH_j)$, a
procedure known as Trotterization~\cite{suzuki1991general,berry2007efficient,poulin2014trotter,babbush2016exponentially,heyl2019quantum,tranter2019ordering,cirstoiu2020variational,bolens2021reinforcement,yao2021adaptive,lin2021real,richter2021simulating,mansuroglu2021variational,keever2022classically,tepaske2022optimal,zhang2023low}.
\begin{figure*}[t]
	\centering
	\includegraphics[width=\linewidth]{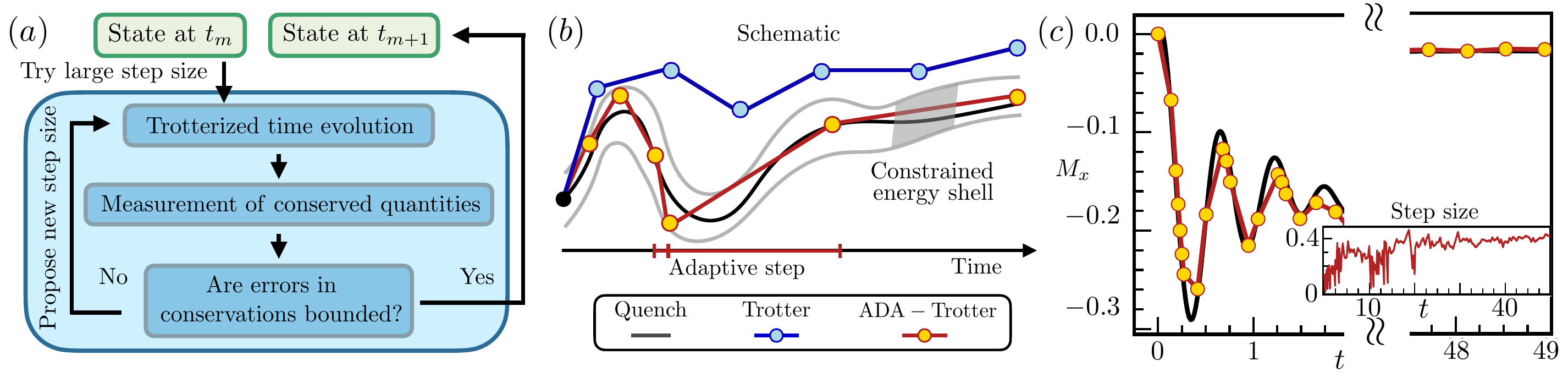}
	\caption{ (a) Schematic of the ADA-Trotter protocol. Given a quantum state at time $t_m$ and a Trotter expansion which approximates a target Hamiltonian, one tries to use the largest possible Trotter step size as long as errors in conservation laws are bounded. (b) ADA-Trotter generally outperforms conventional Trotter: States propagated via ADA-Trotter (orange) are constrained in a energy shell where energy and its variance are approximately conserved. In contrast, a fixed step size is chosen for conventional Trotter (blue) and larger errors may occur. (c) Illustration of local dynamics obtained by ADA-Trotter, which approximates the exact dynamics (black) closely in both short and long times. The inset shows the selected adaptive step size which fluctuates in time. Smaller step sizes are chosen at early times to capture the rapid local oscillations whereas an overall growing trend can be observed at longer times. We use $J_z=-1,h_x=-2,h_z=0.2,L=24$ and initial state polarized in negative $y$ direction for numerical simulation.}
	\label{fig:schematic}
\end{figure*}
The noncommutativity of $H_j$ gives rise to a Trotter error which increases in $\delta t$~\cite{heyl2019quantum};
the longer the simulation time, the larger the error.
Higher-order corrections can be systematically suppressed by using a smaller step size $\delta t$.
However, these come at the price of an increased depth of the gate sequence. Minimization of the circuit depth is key for 
present-day noisy intermediate-scale quantum (NISQ) devices~\cite{preskill2018quantum}, which  experience notable gate imperfections, with efficient quantum error correction still far out of reach~\cite{postler2022demonstration,krinner2022realizing,chen2022calibrated}.
By contrast, unlike in the case of classical computers, the physical (wall-clock) runtime is at present not a similarly significant constraint. It is therefore a natural strategy to focus on minimizing circuit depth, at fixed computation accuracy.

One way to do this is to introduce an adaptive Trotterization step size by exploiting the properties of the evolved state: e.g., at a fixed evolution time, a larger $\delta t$ can give rise to a shorter gate sequence in time windows when the state changes slowly, without incurring a higher error. While this idea underlies adaptive solvers for differential equations and function approximation techniques~\cite{butcher1987numerical}, promoting it to DQS requires addressing two formidable challenges. 
(i) Quantum states are not fully measurable; thus, it is a priori unclear which physical quantity should be used to produce a stable criterion that adapts the step size;
(ii) Trotterization, by replacing continuous with discrete time evolution,  generally violates energy conservation. A variable step size
removes even the remaining discrete time translation invariance, and hence opens further energy absorption channels, manifest in increased approximation errors~\cite{dumitrescu2018logarithmically,else2020long,zhao2021random,long2022many}.

In this work, we propose an adaptive-step Trotterization scheme (ADA-Trotter) for DQS of  many-body dynamics. We focus on time-independent Hamiltonians with arbitrary initial states, i.e., a setting corresponding to a generic quantum quench.
We demonstrate that it outperforms conventional Trotterization while retaining a controllable error in local observables at all simulation times~\footnote{Infidelity in the global wavefunction can also be used to quantify the accuracy of DQS. However, it is not suitable for our purpose because,
	for many-body systems, infidelity significantly overestimates actual errors in local observables}.
An essential element of our algorithm is a feedback loop which self-corrects the simulation errors in the conservation of the target Hamiltonian $H$, through adapting time steps, as summarized in Fig.~\ref{fig:schematic}.
We derive a quantitative estimate of the long-time error for local observables and correlation functions. As a key result, we show that this error is independent of simulation time and the number of qubits. 
While the feedback loop introduces an overhead for each performed time step, the additional resources required depend only polynomially, or logarithmically by using random measurements, on the system size.
At the cost of a moderately increased runtime on the quantum processor, ADA-Trotter opens up the possibility to experimentally implement accurate DQS beyond what is achievable with conventional Trotterization.
\section{Results}
\subsection{Adaptive Trotterization algorithm}
Let us outline the ADA-Trotter algorithm in detail [Fig.~\ref{fig:schematic}(a)].
The key idea behind it is to maximize a variable step size $\delta t_m$ at each time $t_m$, while preserving the expectation value and the variance of the target Hamiltonian $H$, within some predefined fixed tolerances. Below, we show that other constraints, such as the preservation of symmetries, can also be imposed. We illustrate it by preserving gauge invariance, which can critically impact the accuracy of DQS for LGTs.

For a given quantum state $|\psi(t_m)\rangle$, we aim to find the largest possible time step $\delta t_m$, such that, in the time-evolved state $|\psi(t_m{+}\delta t_m)\rangle {=} U_T(\delta t_m) |\psi(t_m)\rangle$,
the energy density 	$\mathcal{E}_{m+1} {=} L^{-1}\langle \psi(t_m+\delta t_m) | H | \psi(t_m+\delta t_m)  \rangle$
and its fluctuations density
$\delta \mathcal{E}^2_{m+1} {=} L^{-1} \langle \psi(t_m+\delta t_m) | H^2 | \psi(t_m+\delta t_m)\rangle {-} L\mathcal{E}^2_{m+1}$ with the system size $L$
remain both bounded:
\begin{equation}
	|\mathcal{E}_{m+1}-\mathcal{E}| < d_\mathcal{E},\qquad \ |\delta \mathcal{E}^2_{m+1}-\delta \mathcal{E}^2| < d_{\delta \mathcal{E}^2},
	\label{eq:e_constraint}
\end{equation}
where $\mathcal{E},\delta\mathcal{E}^2$ are energy and variance density for the initial state.
These conditions ensure the conservation of average density and its fluctuations, up to the maximally allowed errors $d_\mathcal{E}$ and $d_{\delta \mathcal{E}^2}$. Crucially, as we always compare conserved quantities with their initial values, these errors will not accumulate in time.
Remarkably, although we only constrain deviations in the lowest two moments of $H$ explicitly, our numerical results suggest that it is sufficient to constrain the higher moments of $H$.
Hence, the target Hamiltonian $H$ emerges as an approximate constant of motion, despite Trotterization explicitly violating its conservation.
While in conventional Trotterization the error is set by the step size, in ADA-Trotter the error is controlled solely by the tolerances $d_\mathcal{E}$ and $d_{\delta \mathcal{E}^2}$.

\begin{figure*}[t]
	\centering
	\includegraphics[width=\linewidth]{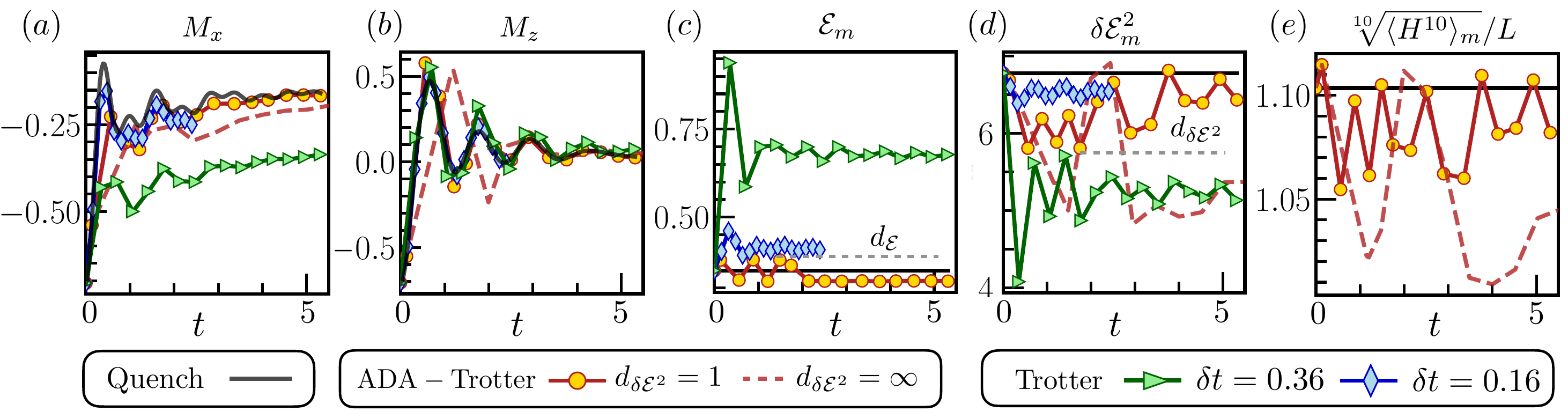}
	\caption{ Comparison between ADA-Trotter and fixed step Trotterization. (a) and (b) depict the time evolution of magnetization in $x$ and $z$ direction, respectively. Variance constraint
		is crucial for
		ADA-Trotter: with variance constraint $d_{\delta \mathcal{E}^2}=1$ (orange circle) 	ADA-Trotter reproduces the exact dynamics whereas notable deviation appears for $d_{\delta \mathcal{E}^2}=\infty$ (dashed red) without variance control. ADA-Trotter outperforms fixed step Trotter: For a large fixed trotter step ($\delta t=0.36$, green triangle), notable local errors appear at early times; A smaller $\delta t$ can be employed (blue diamond) to suppress errors but the total simulation time is limited.  (c) and (d) depict the expectation value of target Hamiltonian $H$ and its variance, deviations of which are constrained by $d_\mathcal{E}$ and $d_{\delta \mathcal{E}^2}$ for ADA-Trotter. Fixed step Trotter can lead to a sufficiently large error in energy at early times.  (e) Higher moments of the target Hamiltonian is better preserved with variance constraint.
		We use $J_z=-1,h_x=-1.7,h_z=0.5,d_\mathcal{E}=0.03,L=24$ for numerical simulation.
	}
	\label{fig:top}
\end{figure*}

The major challenge for ADA-Trotter resides in capturing corrections to the dynamics beyond conventional Trotterization, which we control via a feedback loop that operates as follows [Fig.~\ref{fig:schematic}(a)]:
First, a large time step $\delta t_m$ is chosen.
We implement the time evolution $U_T(\delta t_m)$ on the quantum processor, yielding a candidate state $|\tilde \psi(t_m{+}\delta t_m)\rangle{=}U_T(\delta t_m)|\psi(t_m)\rangle$.
For this candidate state, we measure the energy density
$\tilde{\mathcal{E}}_{m+1}$ and its fluctuations $\delta \tilde{\mathcal{E}}^2_{m+1}$ which can be accessed straightforwardly on quantum computers~\cite{kokail2019self,mi2021information,naldesi2022fermionic}.
In case the measurement outcome violates the conditions of Eq.~\ref{eq:e_constraint}, a new smaller step size $\delta t_m$ is proposed and
this procedure starts over again.

Note that
the state $|\psi(t_m)\rangle$ collapses after projective measurements and hence it needs to be regenerated. However,
once the sequence of the step size before $t_m$ has been determined, the actual runtime for regeneration of the state is fast on most NISQ devices, and determining energy and its variance require only polynomially many local measurements. By using classical shadows with random measurements~\cite{huang2020predicting,huang2021efficient,elben2023randomized}, the measurement cost can be further improved to logarithmic dependence in system sizes, see Sec.~SM 2.5 in Supplementary Material (SM).

An efficient way of finding a new suitable $\delta t_m$ is the bisection method, although other search algorithms can also be employed [cf.~discussion regarding the search algorithm Sec.~SM 2.1 in SM].
Once a suitable $\delta t_m$ has been found, we obtain the quantum state $|\psi(t_m+\delta t_m)\rangle$ at the next time step, and repeat the procedure.
\subsection{Classical emulation}
Let us illustrate the ADA-Trotter algorithm in a classical simulation. Although our algorithm is independent of the underlying model, for concreteness, we consider a non-integrable quantum Ising model, $H=H_++H_-$, as our target Hamiltonian:
\begin{equation}
	H_-= J_z \sum_{j} \sigma_{j}^{z} \sigma_{j+1}^{z}+h_z \sum_{j} \sigma_{j}^{z}, \ \ \    H_+=h_x \sum_{j} \sigma_{j}^{x},
\end{equation}
for a chain of $L$ lattice sites.
We consider a uniform nearest-neighbor Ising coupling $J_z$, and transverse and longitudinal fields $h_x$ and $h_z$, respectively. Periodic boundary conditions are used unless otherwise specified. 
Without loss of generality, we employ a second-order Trotter–Suzuki decomposition,
$
U_T(\delta t) = e^{-i\delta tH_-/2}e^{-i\delta tH_+}e^{-i\delta tH_-/2},
$
to represent the target time evolution operator as a sequence of elementary gates. 

As an example, in Fig.~\ref{fig:schematic}(c) we show the time evolution of the magnetization $M_{x}(t_m) = L^{-1}\sum_j\langle \sigma_j^{x}\rangle_{m}$ for an initial state polarized in the negative $y$ direction; $\langle ...\rangle_{m}$ denotes expectation in the state $|\psi(t_m)\rangle$.  ADA-Trotter (orange circles) closely approximates the exact solution (black). It not only correctly captures the early time oscillations, but also the relaxation of local observables at longer times. Details of the full time evolution and the performance gain compared with conventional Trotterization are discussed in Sec.~SM 2.6. The inset shows the adaptive stepsizes which can fluctuate by one order of magnitude, $\delta t_m\in[0.01,0.46]$, demonstrating the flexibility and the advantage of the adaptive procedure.
The bisection search method makes 10 attempts on average before it can identify the optimal step size. Crucially, the attempt number does not scale up for larger system sizes, suggesting that the search algorithm can also be efficiently implemented on quantum processors with a large number of qubits, see discussions in Sec.~SM 2.3.

We now restrict the simulation to a maximum number of $N{=}15$ Trotter steps; this is equivalent to limiting the circuit depth, and reflects constraints on present-day NISQ devices.
Consider the initial state $\exp(-i\pi\sum_j\sigma_j^y/8)|\downarrow\dots\downarrow\rangle$. The dynamics of the magnetization in the $x$ and $z$ directions are shown in Fig.~\ref{fig:top} (a) and (b), respectively.
At early times, the exact dynamics (black) exhibits rapid oscillations in both observables, which damp out with time.
For ADA-Trotter with tight constraints in energy and variance, $d_\mathcal{E}=0.03,d_{\delta \mathcal{E}^2}=1$ (orange circles), the exact dynamics is reproduced with high accuracy.
We include here both cases of a constrained energy variance with $d_{\delta \mathcal{E}^2}=1$, and of an unconstrained variance $d_{\delta \mathcal{E}^2}=\infty$ (dashed red), to demonstrate the importance of preserving the second moment of energy. 

ADA-Trotter can outperform Trotterization with a fixed step size: To reach the same maximal physical simulation time ($t{\approx}5.5$), we apply Trotterization with a fixed step size $\delta t{=}0.36$ (green triangles), for which notable deviations in $M_x$ appear already at very early times ($t{\approx} 1$); a smaller step size $\delta t{=}0.16$ (blue diamonds) can be chosen to suppress local errors such that they are comparable to ADA-Trotter. However, this comes at a cost of a reduced total simulated time ($t{\approx}2.5$) which drops by a factor of two compared to ADA-Trotter. 

The reason behind the efficiency of
ADA-Trotter with tight constraints lies in the self-correction of errors in the conservation law of the target Hamiltonian, a feature that is crucial for a faithful digital simulation but absent in the conventional Trotter scheme.
As illustrated in Fig.~\ref{fig:top} (c), rather than being a conserved quantity (black), energy density $\mathcal{E}_m$ now becomes time-dependent for ADA-Trotter (orange dots). Remarkably, it fluctuates around the correct value within a controlled error $d_\mathcal{E}$ (grey), signifying that our algorithm is self-correcting the energy errors by adapting the time steps. Similar behavior also occurs to the simulation shown in Fig.~\ref{fig:schematic} (c), see details in Sec.~SM 2.6. In contrast, conventional Trotter (green) leads to notable deviations in energy, which saturate quickly at a value far from the black line and cannot be corrected.

At later times, energy for ADA-Trotter saturates in a preferred direction towards the center of the many-body spectrum ($\mathrm{Tr}[H] /L{=}0$) that corresponds to infinite temperature. This behavior is reminiscent of heating, which commonly occurs in time-dependent quantum chaotic systems where all initial states eventually reach infinite temperature due to energy absorption from the time-dependent drive~\cite{lazarides2014equilibrium,machado2019exponentially,rubio2020floquet}. However, the explicit constraints imposed in ADA-Trotter (Eq.~\ref{eq:e_constraint}) strictly forbid such a heat death. 

Similarly, the energy variance also changes in time and notable deviation from the initial value (black line) can occur if there is no constraint in variance ($d_{\delta \mathcal{E}^2}{=}\infty$, dashed red line in Fig.~\ref{fig:top} (d)). Such a deviation can be controlled by setting $d_{\delta \mathcal{E}^2}{=}1$ (orange circles)~\footnote{By contrast, for fixed-step Trotterization, large deviations in energy and variance appear at very early times ($t{\approx} 1$) for $\delta t{=}0.18$ (green triangle) shown in Fig.~\ref{fig:top} (c) and (d). Therefore, although Floquet theory suggests that the Floquet Hamiltonian $H_F$ defined through the relation $U_T(\delta t){=} \exp(-iH_F\delta t)$ is quasi-conserved~\cite{kuwahara2016floquet,heyl2019quantum}, $H_F$ can still significantly deviate from the target Hamiltonian and induce notable local errors in DQS.}. In Fig.~\ref{fig:top} (e) we show a higher moment of the target Hamiltonian $\sqrt[n]{\langle H^n\rangle_{m}}/L$ for $n=10$: remarkably,
the errors in this quantity are also constrained at $d_{\delta \mathcal{E}^2}{=}1$, and grow with $d_{\delta \mathcal{E}^2}$. 
From this analysis, we conclude that enforcing only energy conservation is insufficient, while constraining in addition its variance is necessary to preserve the conservation of the target Hamiltonian. Similar behavior also occurs at the integrable point ($h_z=0$), see Sec.~SM 2.8. It happens possibly because by central-limit theorem the correct characterization of the energy distribution requires only the first two moments~\cite{hartmann2004gaussian}. In particular, consider a Hamiltonian with nearest neighboring interactions and a product state $\ket{\psi(0)}$ that are accessible on most current digital devices. Suppose this state has mean energy $E_0$ and energy variance $\delta E_0^2$, such that the variance is lower bounded by the number of qubits $\delta E^2\ge aL$ with $a>0$, the energy distribution has been shown to converge to a Gaussian $\rho(E)\sim e^{-\left(E-E_0\right)^2 / 2 \delta E^2}$ in the thermodynamic limit~\cite{hartmann2004gaussian}. The condition can be easily verified as shown in Sec.~SM 2.7. In Sec.~SM 2.8 we also fine-tune the initial state such that it is far from a Gaussian distribution, hence, ADA-Trotter may not perform well. However, we expect central limit theorem should hold true for most generic realistic systems and states. 

It is worth noting that the instantaneous long-time errors for ADA-Trotter can be orders of magnitude smaller compared to the error at short times. For instance, in Fig.~\ref{fig:top} (a) with variance control (orange circles) at $t{=}1.2$, the error in $M_x$ is close to 0.1, which drops to 0.004 at $t{=}5.5$; the same behavior occurs even without variance control (red dashed line).
This is counter-intuitive since Trotter errors in the global wave function are expected to accumulate in time~\cite{tepaske2022optimal}. In the following, we will rationalize this observation with the help of the eigenstate thermalization hypothesis (ETH), and derive a quantitative estimate for the ADA-Trotter errors in local observables.
\subsection{Control of asymptotic errors}
In a standard quench setup, according to ETH, the long-time averaged expectation value of a local observable $O$ can be well-captured by the diagonal ensemble prediction~\cite{rigol2008thermalization} 
\begin{equation}
	\begin{aligned}
		\label{eq.O_DE}
		O_{\mathrm{diag}}(\mathcal{E})\approx O(\mathcal{E})+\frac{\delta \mathcal{E}^2}{2L}{O''}(\mathcal{E}).
	\end{aligned}
\end{equation}
Here $O(\mathcal{E})=\langle \mathcal{E}|O|\mathcal{E}\rangle$ is the  micro-canonical value for eigenstate $|\mathcal{E}\rangle$ at energy density $\mathcal{E}$. According to ETH, $O(\mathcal{E})$ is a smooth function of energy density and ${O''}(\mathcal{E})$ denotes the second derivative w.r.t.~$\mathcal{E}$. Usually, for locally interacting systems the variance density $\delta \mathcal{E}^2$ does not scale with the system size; hence, the second contribution on the right-hand side vanishes in the thermodynamic limit ($L\to\infty$)~\cite{rigol2008thermalization}. 
\begin{figure}[t]
	\centering
	\includegraphics[width=\linewidth]{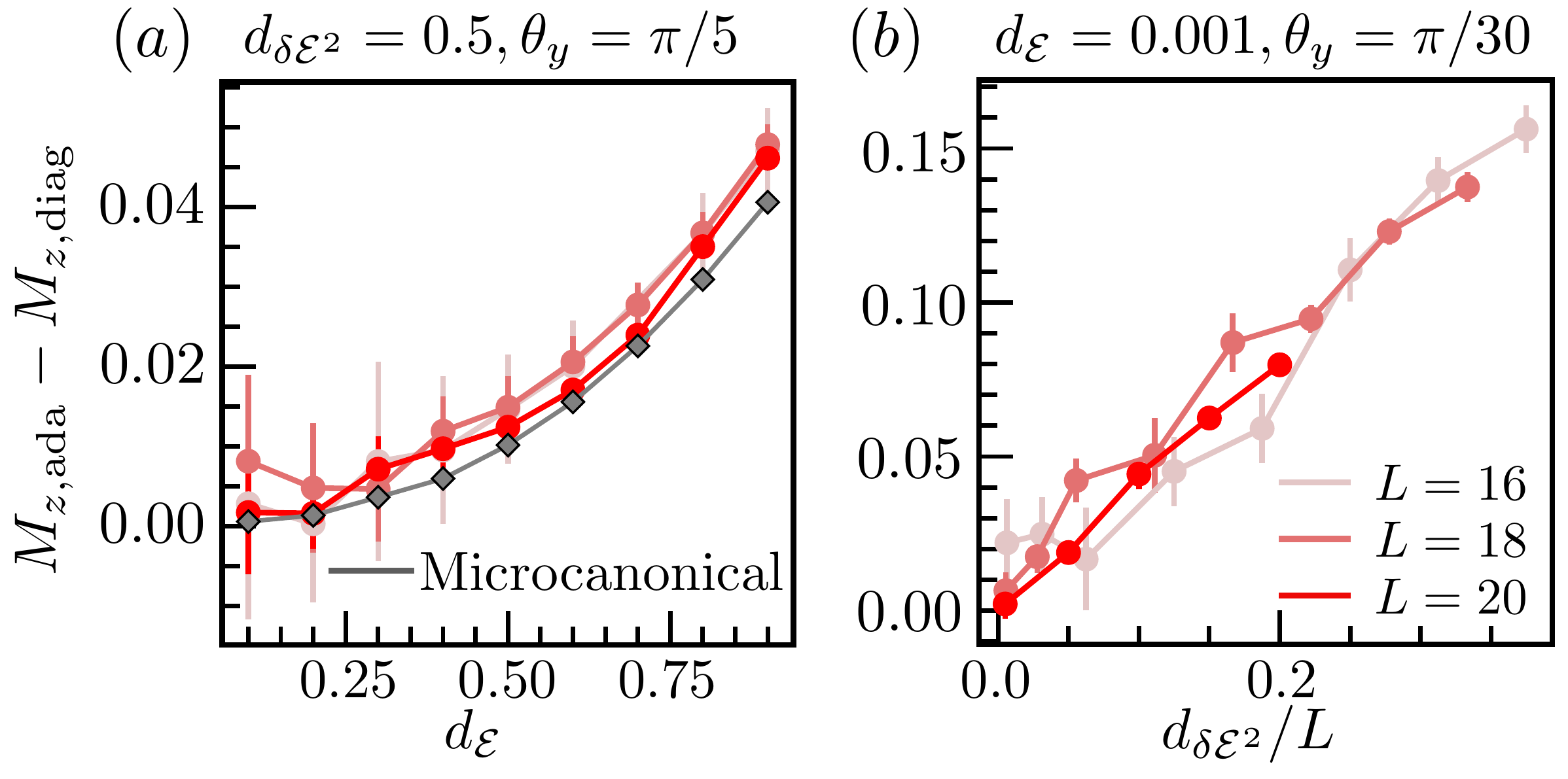}
	\caption{Deviation in the expectation value for the z-magnetization between the ADA-Trotter and exact quench results in the long time limit, as a function of (a) energy density deviation $d_{\mathcal{E}}$ and (b) variance deviation $d_{\delta \mathcal{E}^2}$. (a) The local error decreases for a smaller energy deviation and can be captured by the micro-canonical prediction (grey). (b) Variance deviation leads to local errors as finite size effects, which vanishes linearly in $d_{\delta \mathcal{E}^2}/L$. Parameters $J_z=1,h_x=1,h_z=0.3$ and initial states $\exp(-i\theta_y\sum_j\sigma_j^y)|\downarrow\dots\downarrow\rangle$ are used for numerical simulation.
	}
	\label{fig:z_error}
\end{figure}

For ADA-Trotter, at long times, we assume that the system also reaches a diagonal ensemble at the shifted energy density $\mathcal{E}+ d_{\mathcal{E}}$ and variance density $\delta\mathcal{E}^2+d_{\delta\mathcal{E}^2}$. Then,
one can perturbatively expand $O(\mathcal{E})$ in Eq.~\ref{eq.O_DE} in terms of  $d_{\mathcal{E}}$ to estimate the local error for $L\to\infty$ as
\begin{equation}
	\label{eq.O_AE_d_thermo}
	\begin{aligned}
		{O_{\mathrm{ada}}}
		-O_{\mathrm{diag}}(\mathcal{E})=d_{\mathcal{E}}{O'}(\mathcal{E})+\frac{d_{\mathcal{E}}^2}{2}{O''}(\mathcal{E})+\mathcal{O}(d_{\mathcal{E}}^3),
	\end{aligned}
\end{equation}
where $O_{\mathrm{ada}}{=}O_{\mathrm{diag}}(\mathcal{E}{+}d_{\mathcal{E}})$ denotes the diagonal-ensemble prediction for ADA-Trotter. The explicit derivation is
given in the Sec..
Eq.~\ref{eq.O_AE_d_thermo} suggests a linear dependence on $d_{\mathcal{E}}$ either when $d_{\mathcal{E}}$ is small, or when ${O''}(\mathcal{E})$ vanishes (see Fig.~S18 in the SM). In Fig.~\ref{fig:z_error} (a), we show this error extracted from the long-time average of $M_z$ as a function of $d_{\mathcal{E}}$ for different system sizes. Error bars correspond to the standard deviation of $M_z$ in the large time window used to perform the time average, see details in Sec.~SM 2.10. Clearly, for a smaller $d_{\mathcal{E}}$, the local error decreases and vanishes in the limit $d_{\mathcal{E}}\to 0$ for large systems ($L\geq 18$). We also compute the deviation ${O(\mathcal{E}{+}d_{\mathcal{E}})}
{-}O(\mathcal{E})$ via exact diagonalization of the target Hamiltonian for $L=20$, which matches well with the error in the time evolution in a large range of $d_{\mathcal{E}}$. 

Our results imply that, for DQS of a quenched system in the thermodynamic limit, although variance deviations may introduce errors in ADA-Trotter at early times (as shown in Fig.~\ref{fig:top}), they are suppressed by
quantum thermalization at long times.   A tight constraint in energy conservation suffices to bound errors in local observables asymptotically. This also holds true for correlation functions as long as ETH remains valid. By contrast, conventional Trotterization enters challenging regimes at long times and exhibits uncontrollable heating.

For any finite system size, however, errors in energy variance contribute to local observables. Understanding them is important since many present-day DQS platforms are intermediate-scale. These errors become particularly notable for
$d_{\mathcal{E}}\ll d_{\delta \mathcal{E}^2}/{L}$  where the error is mostly generated by the variance deviation and one obtains
${O_{\mathrm{ada}}}{-}O_{\mathrm{diag}}{\approx}{d_{\delta \mathcal{E}^2}}{O''}(\mathcal{E})/{2L}$, see more general discussions in Appendix \ref{sec.appB}. In Fig.~\ref{fig:z_error} (b), we verify this linear dependence on $d_{\delta \mathcal{E}^2}/L$ in the asymptotic error on $M_z$ for different $d_{\delta \mathcal{E}^2}$ and system sizes. Since it tends to cross the origin, this error becomes negligible for sufficiently large system sizes. We would also like to mention that, before reaching the tolerance in energy, a tight variance bound induces non-trivial transient dynamics with a constrained heating rate, see Sec.~SM 2.9. 
\subsection{Protection of gauge symmetry}
One crucial generalization of ADA-Trotter is to explicitly impose extra constraints in the feedback loop in addition to Eq.~\ref{eq:e_constraint}, such that other desired constants of motion can be preserved. This is impossible within the conventional Trotter scheme and independent of underlying models, highlighting the potential of ADA-Trotter in DQS where symmetry is playing an important role. To illustrate this feature, we focus on protecting the local Gauss's law which is crucial for LGTs, yet can be easily lost in real quantum simulators~\cite{schweizer2019floquet,yang2020observation}. 

Consider the paradigmatic spin-$S$ $U(1)$ quantum link model in (1+1) dimensions, which is commonly used as a lattice version of quantum electrodynamics
~\cite{mildenberger2022probing}. The model has two species of particles and can be described by the Hamiltonian  ${H}= H_{\mathrm{kin}}+H_{\mathrm{free}}$ with
\begin{eqnarray}
	\begin{aligned}
		&H_{\mathrm{kin}} =  \sum_{j}\frac{J}{ 2\sqrt{S(S+1)}}\left({\sigma}_{j}^{+} {s}_{j, j+1}^{+} {\sigma}_{j+1}^{-}+\text {h.c. }\right),\\
		&H_{\mathrm{free}}= \sum_{j}{\mu}(-1)^j {\sigma}_{j}^{z}+{k}\left({s}_{j, j+1}^{z}\right)^{2}.
	\end{aligned}
\end{eqnarray}
Here the spin-$1/2$ Pauli operator $\sigma_j^{\pm}$ creates or annihilates the matter field on site $j$; the $s_{j,j+1}$ operator represents the spin-$S$ degree of freedom (d.o.f.) positioned at the links between sites $j$ and $j+1$; $J$ is the kinetic energy term which couples matter and gauge fields; $\mu$ denotes the bare matter mass, and $k$ is the electric-field coupling strength. We consider $L$ matter sites and periodic boundary conditions. Gauss's law implies a $U(1)$ gauge symmetry generated by the operator 
${G}_{j}=\left[\sigma^z_{j}+{s}_{j-1, j}^{z}-{s}_{j, j+1}^{z}+(-1)^{j}\right]/2$, which commutes with the Hamiltonian $H$ for any $j$. Therefore, the Hilbert space separates into exponentially many disconnected symmetry sectors. In the following, we focus on the sector satisfying Gauss's law, $G_j|\psi(0)\rangle{=}0$, for any $j$, and use spin-1 gauge fields. Our method also applies for general spin-$S$ d.o.f.~and other symmetry sectors as discussed in Sec.~SM 1.5 in the SM.

We consider the initial state $|\psi(0)\rangle =   |\uparrow\downarrow\dots\uparrow\downarrow\rangle_{\sigma}\otimes|0\dots0\rangle_{s}$ fulfilling Gauss's law. To simulate the Trotterized time evolution, we consider for concreteness the decomposition
$H_+= H_{\mathrm{kin}}+\lambda V,\ \     H_-=H_{\mathrm{free}}-\lambda V$
with a gauge-breaking perturbation $V$. The average of $H_+$ and $H_-$ reproduces the target Hamiltonian $H$. For zero perturbation strength $\lambda$, Gauss's law is ensured for all simulation times; for non-zero $\lambda$, the expectation value of the gauge generator $\mathcal{G}_m(j)=\langle G_j\rangle_{m}$ on site $j$, and its variance $\delta \mathcal{G}_m^2(j)=\langle G_j^2 \rangle_{m} - \mathcal{G}_m^2(j)$ for the state $|\psi (t_m) \rangle$ obtained by the ADA-Trotter at time $t_m$, become time-dependent and non-zero. To preserve the gauge symmetry, in addition to Eq.~\ref{eq:e_constraint}, we impose the conditions
$
\sum_j|\mathcal{G}_{m}(j)|/L < d_\mathcal{G},\sum_j |\delta \mathcal{G}^2_{m}(j)|/L < d_{\delta \mathcal{G}^2} ,
$ in the feedback loop to bound the density of gauge violation. 

\begin{figure}
	\centering
	\includegraphics[width=\linewidth]{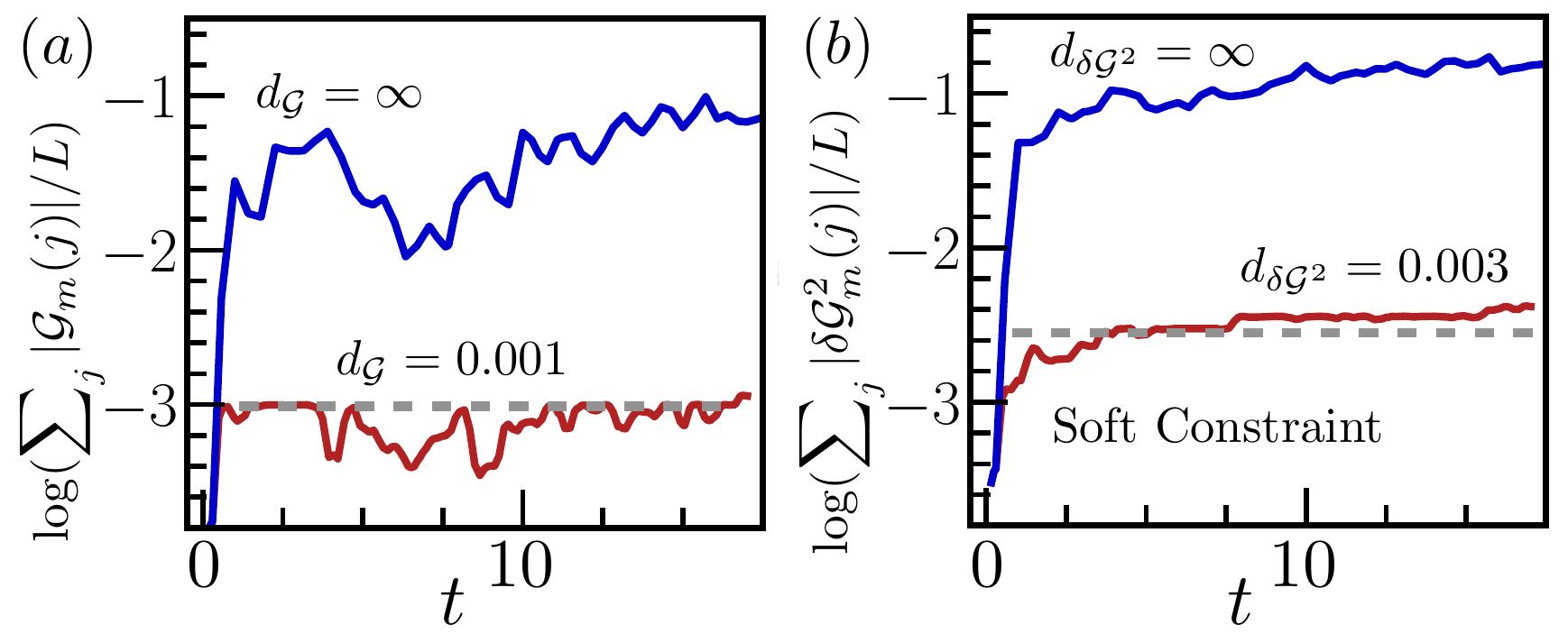}
	\caption{Dynamics of the expectation values of the gauge symmetry generator in (a) and its variance in (b). Additional constraints on gauge violation preserve the gauge symmetry (red), otherwise errors quickly grow in a short time (blue). A soft constraint is used to prevent the ADA-Trotter from freezing. We use $J=0.5,\mu=0.5,k=0.5,d_{\mathcal{E}}=0.1,d_{\delta \mathcal{E}^2}=0.2,\lambda=0.3, L=6$ and
		$
		V = \sum_j\left[{{s}_{j, j+1}^{+}}/{ \sqrt{S(S+1)}}+\sigma_j^{+}\sigma_{j+1}^-+\text {h.c.}\right]
		$ for numerical simulation. }
	\label{fig:symmetry_protection}
\end{figure}
In Fig.~\ref{fig:symmetry_protection}, we plot the errors in the expectation value of the gauge generator and its variance on a log-scale in panels (a) and (b), respectively. As the system involves both the matter and gauge d.o.f., the numerical simulation is limited to a small system size $L=6$. Without constraints in gauge symmetry (blue line, $d_\mathcal{G}=\infty,d_{\delta \mathcal{G}^2}=\infty$), the errors in both panels quickly increase at short times and saturate around 0.1, indicating a severe violation of Gauss's law. By contrast, the additional constraint $d_\mathcal{G}=0.001,d_{\delta \mathcal{G}^2}=0.003$ (red line) significantly bounds the errors by two orders of magnitude, and hence the gauge symmetry is controllably preserved. 

The quick saturation of the errors measuring the violation of Gauss's law to a predefined tolerance (grey line), can cause ADA-Trotter to `freeze' (i.e., the algorithm tends to always choose the smallest possible step size). Consequently, the quantum state barely propagates, despite the large number of quantum gates consumed. Freezing also occurs in quantum Ising models if energy constraints are too tight. To resolve this issue, rather than using fixed tolerances, here we employ a soft constraint such that tolerances can increase by $30\%$ of their original values whenever the smallest step size is chosen, see SM for more details.	
\section{Discussion}
The ADA-Trotter algorithm we propose is capable of simulating local time evolution for DQS of time-independent systems with controllable errors at all times. This adaptive and self-correcting scheme can be particularly useful when there are different timescales throughout the evolution, which typically happens as a result of quantum thermalization separating the early-time coherent oscillations and long-time relaxation. We find that quantum thermalization furthermore suppresses errors in local observables and ensures the long-time stability of ADA-Trotter, which is absent in conventional Trotter where systems normally exhibit uncontrollable heating to infinite temperature. 
It would be worth benchmarking ADA-Trotter when thermalization is absent, for instance, when systems are integrable~\cite{essler2016quench}, long-range interacting~\cite{sugimoto2022eigenstate}, many-body localized~\cite{abanin2019colloquium} or host quantum many-body scars~\cite{serbyn2021quantum}. Those non-thermalizing systems normally have an extensive number of conserved quantities, for which extra constraints in the feedback loop may be required.

ADA-Trotter remains robust and can still self-correct errors in energy even in the presence of dissipation, as long as the noise strength is weak, see Sec. SM 1.2 for more details. It would also be interesting to use error mitigate to further improve the accuracy of this algorithm~\cite{endo2019mitigating,yang2021accelerated,zhang2022unbiased,takagi2022fundamental,yang2023simulating}.

We generalize ADA-Trotter to protect symmetries in addition to the energy constraints and we demonstrate this by controlling the violation of Gauss's law in U(1) LGTs. By preserving the time evolution in a given symmetry sector, ETH still applies and ADA-Trotter remains stable at long times. Such symmetry protection is independent of the microscopic details of the target systems, hence, we expect it to have potentially a wide range of applicability.
Note that such symmetry protection preserves the unitarity of the time evolution; hence, it is fundamentally different from post-selection which may introduce additional errors in DQS~\cite{vovrosh2021confinement,nguyen2022digital}.

In practice, more sophisticated classical optimization routines may further improve the efficiency of ADA-Trotter. For instance, Bayesian optimization can minimize the experimental efforts in measurement~\cite{sauvage2020optimal}, and reinforcement learning may be useful in determining optimal step sizes and avoiding freezing~\cite{bukov2018reinforcement,niu2019universal,wauters2020reinforcement,bolens2021reinforcement}. 

Beyond DQS of quenched problems, Trotterization has been employed in other contexts, e.g., quantum imaginary time evolution~\cite{motta2020determining}, time-dependent Hamiltonian simulation~\cite{poulin2011quantum,lau2021noisy,watkins2022time} and classical numerical algorithms like the time-evolving block decimation method~\cite{schollwock2011density}. Our control scheme paves pathways to making these algorithms adaptive in time, and improve their performance. 
\\
\textbf{Acknowledgments}
We thank Anatoli Polkovnikov, Johannes Knolle, Paul Schindler, Andrea Pizzi, Joseph Vovorsh, Markus Drescher for enlightening discussions. This work is in part supported by the Deutsche Forschungsgemeinschaft  under  cluster of excellence ct.qmat (EXC 2147, project-id 390858490).
M.B.~was supported by the Marie Sk\l{}odowska-Curie grant agreement No 890711.
This project has received funding from the European Research Council (ERC)
under the European Unions Horizon 2020 research and innovation programme
(grant agreement No. 853443).
This research was partially supported by the ARC DP210101367.\\
\textbf{Author contributions:} M.H. proposed the research. H.Z. proposed the algorithm and performed the analytical calculations and numerical simulations. All authors discussed the results and wrote the manuscript.\\ \textbf{Competing interests:} The authors declare that they have no competing interests.\\
\textbf{Data and materials availability:} All data needed to evaluate the conclusions in the paper are present in the paper and/or the Supplementary Materials. Additional data and code available on GitHub (https://github.com/ZhaoHongzheng/ADA-Trotter) or from first author upon request. 
\appendix
\section{Numerical methods}
\label{sec.appA}
We use the open-source Python package QuSpin to perform exact diagonalization and quantum dynamics~\cite{weinberg2017quspin}. We  employ a second-order Trotter–Suzuki decomposition,
$
U_T(\delta t) = e^{-i\delta tH_-/2}e^{-i\delta tH_+}e^{-i\delta tH_-/2},
$ to realize the trotterized time evolution. Numerically, the exp$\_$op class in QuSpin is used, which does not calculate the actual matrix exponential of $U_T$, but instead computes the action of the matrix exponential through its Taylor series. For translation invariant systems, we restrict the basis in the zero momentum block to enable the simulation of larger system sizes. If the system is parity symmetric with respect to the middle of the chain, we only use basis in the positive parity sector. For the microcanonical prediction of local observables as shown in Fig.~\ref{fig:z_error}, we perform exact diagonalization of the target Hamiltonian also in the zero momentum and positive parity symmetry sector.

To determine the optimal Trotter step size $\delta t$ for ADA-Trotter, different search algorithms (bisection and sequential search) are used, see Sec.~SM 2.1 for more details. Bisection search is used to generate the results in Figs.~\ref{fig:schematic} and \ref{fig:top}, and sequential search is used for the other figures in the main text. 
\section{Long-time stability of ADA-Trotter}
\label{sec.appB}
Based on eigenstate thermalization hypothesis (ETH), we are able to estimate the long-time deviation in expectation values of local observables, for two diagonal ensembles with different energy and energy variance. Both the energy and its variance are controllable with ADA-Trotter.
Consider a generic quantum many-body system satisfying ETH, given an initial state with mean and variance of energy
\begin{equation}
	E = \langle H\rangle_{0}, \ \ \ 
	\delta {E}^2 = \langle H^2\rangle_{0}-\langle H\rangle^2_{0}.
\end{equation}
As the Hamiltonian is time-independent, both quantities will be conserved.
For normal short-ranged interacting system, we have the following scaling~\cite{rigol2008thermalization} 
\begin{equation}
	{E}\sim L,\quad \delta {E}^2 \sim {L},
\end{equation}
where $L$ represents the system size in one dimension. The following results can also be generalized to higher dimensions.  For a local observable $\hat{O}$ we define
$\Bar{O}$ as the long time average
\begin{equation}
	\bar{O} =  \frac{1}{\tau}\int_0^\tau dt \langle \hat{O}\rangle_t,
\end{equation}
where $\langle ...\rangle_{t}$ denotes expectation in the state $\ket{\psi(t)}$. $\bar{O}$ can be captured by the diagonal ensemble prediction as $\bar{O} = O_{\mathrm{diag}}=\sum_m |c_m|^2O_{mm}$ where the diagonal elements of the operator read as
$O_{mm}=\langle m|O|m\rangle$, $c_m=\langle \psi(0)|m\rangle$ for initial state $|\psi(0)\rangle$. By assuming that $O_{mm}$ is a continuous function which can be approximated by its thermal prediction $O(E)$, one can perform a perturbative expansion around the micro-canonical prediction for the diagonal ensemble as~\cite{rigol2008thermalization}
\begin{eqnarray}
	\begin{aligned}
		\label{eq.O_DE_expansion}
		O_{\mathrm{diag}} = \sum_m |c_m|^2\Big[O({E})+(E_m-E){O'(E)}+\frac{1}{2}(E_m\\-{E})^2{O''(E)}+\dots\Big]
		=O({E})+\frac{1}{2}(\delta {E})^2{O''(E)}+\dots,
	\end{aligned}
\end{eqnarray}
where the dependence on $O'(E)$ vanishes. Since energy is an extensive quantity, we can use the density of the energy $\mathcal{E}={E}/L$, so the equation above becomes
\begin{equation}
	\begin{aligned}
		\label{eq.O_diag_SM}
		O_{\mathrm{diag}}
		=O(\mathcal{E})+\frac{1}{2} \frac{\delta E^2}{L^2}{O''(\mathcal{E})}+\dots.
	\end{aligned}
\end{equation}
In the main text, we define the density of energy variance as $\delta \mathcal{E}^2=\delta E^2/L$, hence we obtain $O_{\mathrm{diag}}
=O(\mathcal{E})+{\delta \mathcal{E}^2}{O''(\mathcal{E})}/{2L}+\dots$ which leads to Eq.~\ref{eq.O_DE} of the main text.
For Eq.~\ref{eq.O_diag_SM}, as long as 
\begin{equation}
	\frac{1}{2} \frac{\delta E^2}{L^2}{O''(\mathcal{E})}\ll O(\mathcal{E}),
\end{equation}
one can approximate $O_{\mathrm{diag}}$ solely by $O(\mathcal{E})$.
This is normally the case since $\delta {E}^2$ scales linearly in system size $L$ by making the assumption that there is no long-range connected correlations in the system \cite{rigol2008thermalization}. Thus, $\delta {E}^2/L^2\sim 1/L$ which vanishes in the thermodynamic limit and the diagonal ensemble prediction matches with the canonical prediction as $O_{\mathrm{diag}}(\mathcal{E})
=O(\mathcal{E})$. 

For ADA-Trotter, suppose that the total energy reaches a tolerance bound 
\begin{equation}
	\label{eq.deviation}
	\widetilde{E}={E}+\Delta_{E},\quad {\delta \widetilde{E}}^2=\delta E^2+\Delta_{\delta E^2}.
\end{equation}
For the long-time relaxation,
one can make a similar calculation of $O_{\mathrm{ada}}$ as the diagonal ensemble obtained by ADA-Trotter at shifted energy and variance:
\begin{equation}
	\label{eq.O_AE_pre}
	\begin{aligned}
		O_{\mathrm{ada}}=
		O(\widetilde{E})+\frac{1}{2}(\delta {\widetilde{E}})^2{O''(\widetilde{E})}+\dots,
	\end{aligned}
\end{equation}

We can now Taylor expand Eq.~\ref{eq.O_AE_pre} via Eq.~\ref{eq.deviation}:
\begin{eqnarray}
	\label{eq.O_AE}
	\begin{aligned}
		{O_{\mathrm{ada}}} =
		&O({E})+\Delta_{{E}}{O'}({E})+\frac{\Delta_{{E}}^2}{2}{O''}({E})+\frac{1}{2}\left[{\delta {E}}^2+\Delta_{\delta {E}^2}\right]\\
		\times&\left[{O''}({E})+\Delta_{{E}}{O'''}({E})\right]+\dots\\
		=&\left( O({E})+\frac{1}{2}{\delta {E}}^2{O''}({E})\right)+\Delta_{E}{O'}({E})\\
		+&\frac{\Delta_E^2+\Delta_{\delta E^2}}{2}{O''}({E})+\frac{\delta {E}^2+\Delta_{\delta E^2}}{2}\Delta_E{O'''}({E})+\dots,
	\end{aligned}
\end{eqnarray}
and the first two terms match with $O_{\mathrm{diag}}$ in Eq.~\ref{eq.O_DE_expansion}. One can now express the above equation using energy density 
\begin{eqnarray}
	\label{eq.O_AE_thermodynamic}
	\begin{aligned}
		{O_{\mathrm{ada}}}
		&=O_{\mathrm{diag}}(\mathcal{E})+\frac{\Delta_E}{L}{O'}(\mathcal{E})+\frac{\Delta_E^2+\Delta_{\delta E^2}}{2L^2}{O''}(\mathcal{E})\\
		&+\frac{\delta {E}^2+\Delta_{\delta E^2}}{2L^3}\Delta_E{O'''}(\mathcal{E})+\dots.
	\end{aligned}
\end{eqnarray}
The tolerances, $\Delta_{E,\delta E^2}$, in ADA-Trotter are tunable parameters. Therefore, as long as these prefactors in the above higher order corrections drop to zero for $L\to\infty$, one can guarantee that the long time relexed local observable matches with the exact result $O_{\mathrm{diag}}(\mathcal{E})$. For instance, $\Delta_{E,\delta E^2}$ can be chosen to be a small constant value which does not scale with system size. However, for large system sizes, the ADA-Trotter algorithm may freeze and the system barely propagates, which is inefficient as many quantum gates are wasted. Instead, in our numerical simulation, we consider the density constraints $\Delta_E/L = d_{\mathcal{E}},\quad \Delta_{\delta E^2}/L = d_{\delta \mathcal{E}^2}$ with constants $d_{\mathcal{E},\delta\mathcal{E}^2}$. Eq.~\ref{eq.O_AE_thermodynamic} thus reduces to
\begin{eqnarray}
	\label{eq.O_AE_d}
	\begin{aligned}
		{O_{\mathrm{ada}}}
		&=O_{\mathrm{diag}}(\mathcal{E})+d_{\mathcal{E}}{O'}(\mathcal{E})+\frac{d_{\mathcal{E}}^2L^2+d_{\delta\mathcal{E}^2}L}{2L^2}{O''}(\mathcal{E})\\
		&+\frac{\delta {E}^2+d_{\delta\mathcal{E}^2}L}{2L^2}d_{\mathcal{E}}{O'''}(\mathcal{E})+\dots.
	\end{aligned}
\end{eqnarray}
In the thermodynamic limit in one dimension, one obtains
\begin{equation}
	\begin{aligned}
		\lim_{L\to\infty} {O_{\mathrm{ada}}}
		&=O_{\mathrm{diag}}(\mathcal{E})+d_{\mathcal{E}}{O'}(\mathcal{E})+\frac{d_{\mathcal{E}}^2}{2}{O''}(\mathcal{E})+\mathcal{O}(d_{\mathcal{E}}^3),
	\end{aligned}
\end{equation}
leading to Eq.~\ref{eq.O_AE_d_thermo} in the main text.
It suggests that the long-time error in local observables in ADA-Trotter is dominated by energy deviation. The variance deviation can be ignored in the thermodynamic limit if ETH holds true.
Of course, it is still important to discuss its effects on finite size systems. Suppose we have a small deviation in energy satisfying $d_{\mathcal{E}}\ll {d_{\delta \mathcal{E}^2}}/{L}$. Then, Eq.~\ref{eq.O_AE_d} reduces to
\begin{equation}
	\label{eq.O_AE_d_smalld1}
	\begin{aligned}
		{O_{\mathrm{ada}}}
		&=O_{\mathrm{diag}}(\mathcal{E})+\frac{d_{\delta \mathcal{E}^2}}{2L}{O''}(\mathcal{E}),
	\end{aligned}
\end{equation}
indicating a linear dependence on $d_{\delta \mathcal{E}^2}/L$ in the local error which is verified in Fig.~\ref{fig:z_error} in the main text.

\bibliography{Trotter}
\let\addcontentsline\oldaddcontentsline % Restore \addcontentsline

	\cleardoublepage
	\onecolumngrid
	
	\begin{center}
		\textbf{\large{\textit{Supplementary Material} \\ \smallskip
				Making Trotterization adaptive for NISQ devices and beyond}}\\
		\hfill \break
		\smallskip
	\end{center}
	
	\renewcommand{\thefigure}{S\arabic{figure}}
	\setcounter{figure}{0}
	\renewcommand{\theequation}{S.\arabic{equation}}
	\setcounter{equation}{0}
	\renewcommand{\thesection}{SM\;\arabic{section}}
	\setcounter{section}{0}
	\tableofcontents
	
\section{Additional simulation with different initial states and physical models}

\subsection{Performance of ADA-Trotter against gate imperfections}
\label{sec.random_coupling}
\begin{figure}[h]
	\centering
	\includegraphics[width=0.7\linewidth]{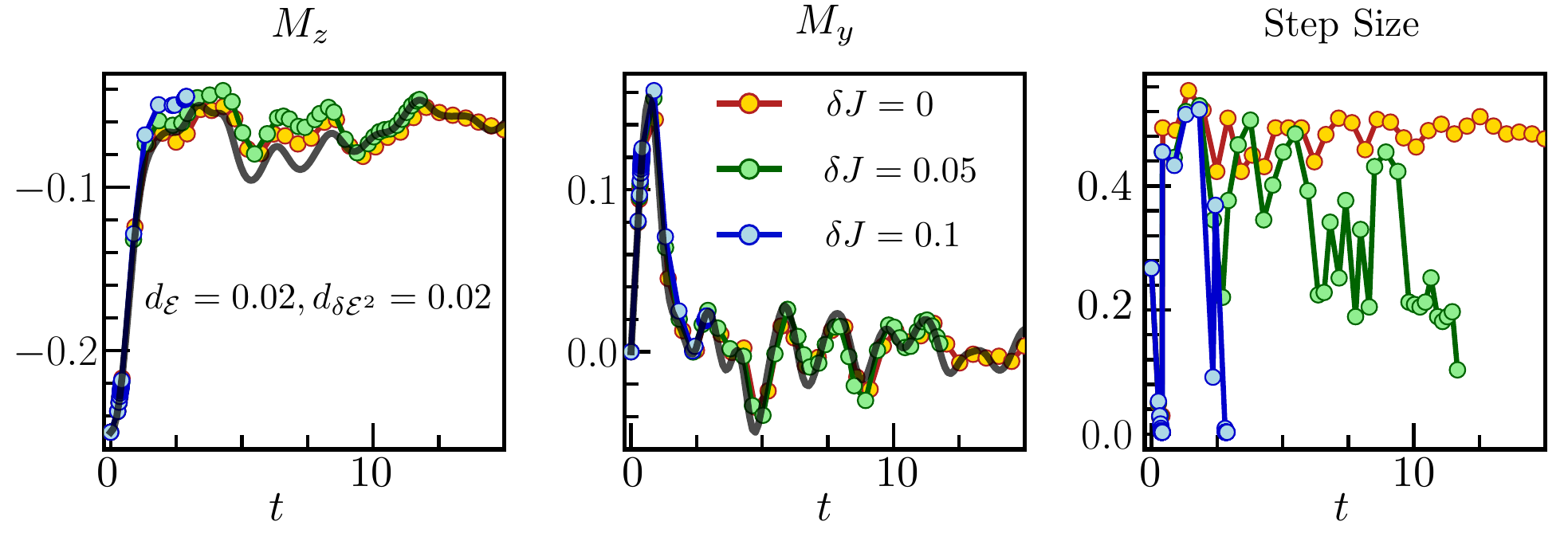}
	\caption{\textbf{[Quantum Ising model with random Ising couplings]} Dynamics of local observables obtained by ADA-Trotter (circles) and exact quench (black line). 
		Disorder in the Ising coupling is introduced to simulate gate imperfections in real devices.
		ADA-Trotter performs well when these imperfections are weak (orange, green). Large imperfection (large $\delta J$) freezes the algorithm, namely ADA-Trotter tends to choose the smallest possible step size and the system barely propagates (blue). We use $J_z=1,h_x=0.6,h_z=0.8,L=16$ and bisection search for the numerical simulation.
	}
	\label{fig:gate_random}
\end{figure}
On real quantum devices, gate imperfections are common which may introduce errors for DQS in addition to Trotter errors. To emulate their effect, 
in this section, we consider adding spatial randomness on top of the uniform Ising coupling. Therefore, now the Trotterized Hamiltonian reads as 
\begin{equation}
	H_+=  \sum_{j} J_j^z\sigma_{j}^{z} \sigma_{j+1}^{z}+h_z \sum_{j} \sigma_{j}^{z}, \ \ \    H_-=h_x \sum_{j} \sigma_{j}^{x},
\end{equation}
with random Ising couplings $J_j^z$ uniformly distributed within the range  $[J_z-\delta J,J_z+\delta J]$. The target Hamiltonian remains the clean model without disorder. We plot the time evolution for three different values of $\delta J$ in Fig.~\ref{fig:gate_random}. Without disorder (orange), ADA-Trotter only exhibits small deviation from the exact solution. For small $\delta J=0.05$ (green), the resulting evolution still captures the quenched dynamics well. The only issue is that the selected step size becomes smaller at later times, indicating that more gate resources are needed than in the clean case. For a relatively large Ising disorder $\delta J=0.1$ (blue), the algorithm freezes and the system barely propagates in time, hence ADA-Trotter fails. One possible solution is to use a soft constraint to bound the error similar to the LGT protection in the main text.

\subsection{Performance of ADA-Trotter with decoherence}
\label{sec:dissipation}
Rather than a single disorder realization with imperfect Ising couplings, we now investigate dissipative effects. We will show that ADA-Trotter can still outperform fixed-step Trotter even in the presence of generic dissipation, as long as the system-environment coupling is weak. 
\begin{figure}[h]
	\centering
	\includegraphics[width=0.75\linewidth]{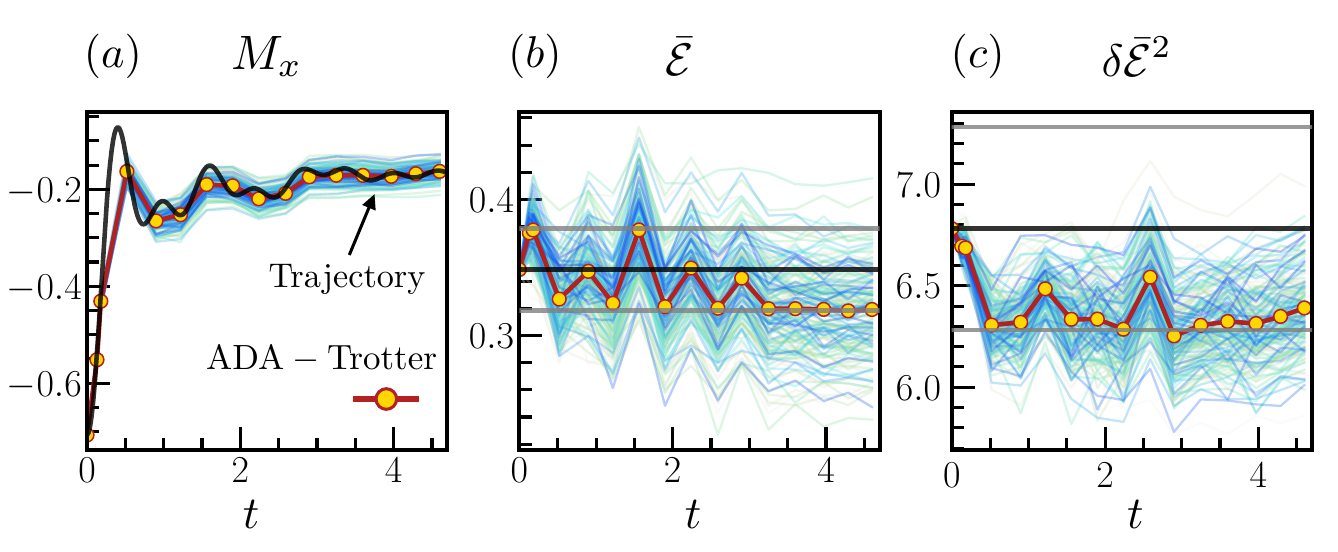}
	\caption{\textbf{[Quantum Ising model with decoherence]} Dynamics of local observables, energy and energy variance for each trajectory (blue lines) and the ensemble (orange dots). Although individual trajectory may not satisfy energy and variance constraints (grey), the ensemble average leads to reliable dynamics. We use the initial state $\exp(-i\pi\sum_j\sigma_j^y/8)|\downarrow\dots\downarrow\rangle$, Hamiltonian parameters $J_z=1,h_x=-1.7,h_z=0.5,\gamma=0.2, L=14$ and $d_{\mathcal{E}}=0.03,d_{\delta\mathcal{E}^2}=0.5$, and total trajectory number $s_{\mathrm{max}}=200$ for numerical simulation.}
	\label{fig:Trajectory}
\end{figure}

The concrete form of dissipation can be generically different for different experimental setups. Therefore, rather than trying to precisely model dissipation in a specific physical platform, we aim to show that our algorithm can function
	well even in the presence of generic weak decoherence process.
	The density matrix of the digital simulator embedded in an environment, can be generally captured by a master equation
\begin{eqnarray}
	\label{eq.dissipation}
	\frac{d}{d t} \rho(t)=-{i}\left[{H}_T(t), \rho(t)\right]+\mathcal{D}[\rho(t)].
\end{eqnarray}
where $H_T$ represents the trotterized Hamiltonian and the $\mathcal{D}[\cdot]$ induces non-unitary dissipation, resulting from the interaction between the system and environment. Simulating Eq.~\ref{eq.dissipation} directly can be numerically costly, especially for many-body systems that are most relevant to our study. Therefore, instead of using the master equation, we consider a stochastic process generated by Hamiltonian $H_s(t)$, that is similar to $H_T(t)$ but involves spatial-temporal randomness, to simulate the non-unitary time evolution. 

We still fix the Trotter decomposition as in the main text, but now we
introduce spatial-temporal randomness as
\begin{eqnarray}
	\begin{aligned}
		&U_T(t,\delta t,s)= e^{-i\delta tH_-/2}e^{-i\delta tH_+}e^{-i\delta tH_-/2}, \ \mathrm{with} \\
		&H_-(t,s)=  \sum_{j} J^z_{j}(t,s) \sigma_{j}^{z} \sigma_{j+1}^{z}+\sum_{j} h_j^z(t,s)\sigma_{j}^{z},\ \  H_+(t,s)= \sum_{j} h_j^x(t,s)\sigma_{j}^{x},
	\end{aligned}
\end{eqnarray}
where $J_{j}(t,s)=J_z+\delta J^z_j(t,s),h^z_{j}(t,s)=h_z+\delta h^z_j(t,s),h^x_{j}(t,s)=h_x+\delta h^x_j(t,s)$, with $J^z_j,\delta h^z_j,\delta h^x_j$ uniformly distributing within the range $\gamma[-J_z,J_z],\gamma[-h_z,h_z],$ and $\gamma[-h_x,h_x]$, respectively. $\gamma$ is a controllable parameter to quantify the system-environment coupling: $\gamma=0$ represents coherent Trotterization dynamics, and larger $\gamma$ generally leads to stronger dissipation. For numerical simplicity, randomness is fixed during each Trotter step, e.g., $\delta J_j(t,s)$ is constant for $t\in(t_m,t_m+\delta t_m)$, but randomly chosen for different position $j$ and time $t_m$, as well as stochastic process $s$. Although the system is initialized in a pure state, numerically we can still generate a classical ensemble of $s_{\mathrm{max}}$ trajectories as $\{\ket{\psi_s(0)}\}$. For each trajectory, it evolves with random Trotterized evolution operator, leading to a new ensemble at time $\delta t$ as $\{\ket{\psi_s(\delta t)}\}=\{U_T(t,\delta t,s)\ket{\psi_s(0)}\}$.
	By averaging over different realizations one can obtain the noise-averaged energy density 	$\bar{\mathcal{E}} {=} s_{\mathrm{max}}^{-1}L^{-1}\sum_s \langle \psi_s(\delta t) | H | \psi_s(\delta t) \rangle$
	and its fluctuations density
	$\delta \bar{\mathcal{E}}^2{=} s_{\mathrm{max}}^{-1}L^{-1}\sum_s( \langle \psi_s(\delta t) | H^2 | \psi_s(\delta t)\rangle {-} \langle \psi_s(\delta t) | H | \psi_s(\delta t) \rangle^2)$. We search for the largest possible Trotter step size $\delta t$ such that Eq.~\ref{eq:e_constraint} is satisfied.
	This process will be repeated to propagate the ensemble for longer times $\{\ket{\psi_s(t_m+\delta t_m)}\}=\{U_T(t,\delta t,s)\ket{\psi_s(t_m)}\}$.

\begin{figure}[h]
	\centering
	\includegraphics[width=0.95\linewidth]{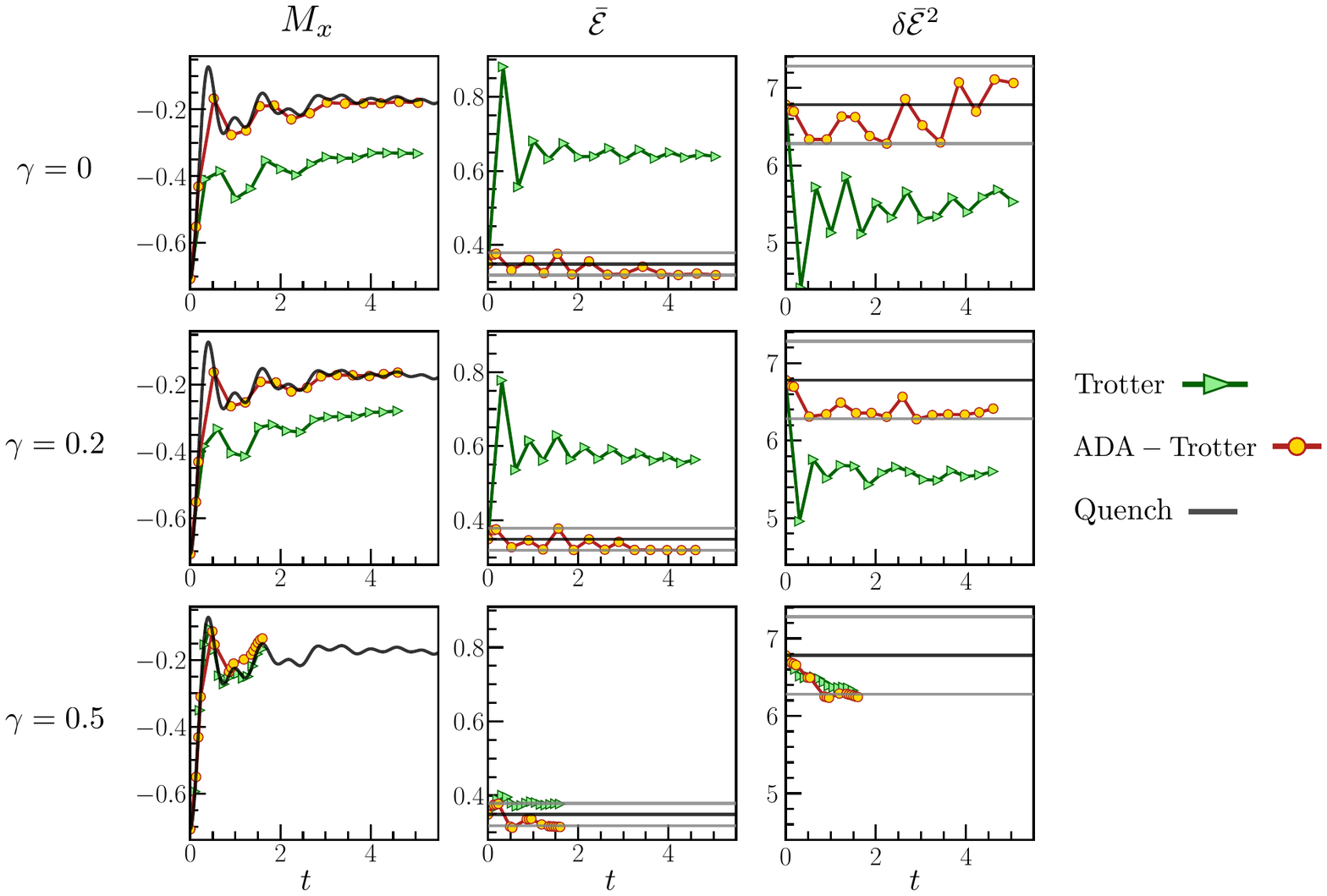}
	\caption{\textbf{[Quantum Ising model with decoherence]} Dynamics of local observables, energy and energy variance. ADA-Trotter still works well in the presence of weak dissipation ($\gamma=0.2$) and outperforms Trotter algorithm. For large dissipation ($\gamma=0.5$), the algorithm tends to freeze because no suitable step size can be identified to satisfy energy constraints (grey lines). We use $J_z=1,h_x=-1.7,h_z=0.5,L=14$ and $d_{\mathcal{E}}=0.03,d_{\delta\mathcal{E}^2}=0.5$ for numerical simulation.}
	\label{fig:Noisy}
\end{figure}
Fig.~\ref{fig:Trajectory} illustrates the dynamics of $s_{\mathrm{max}}=200$ trajectories (blue line) and their ensemble average (orange dots) for $\gamma=0.2$. Clearly, extra spatio-temporal randomness generally induce more errors in DQS, and 
	individual trajectory may violate the constraints in energy and variance. However, ADA-Trotter algorithm can still self-correct errors (including errors generated from both Trotterization and dissipation) in energy and variance for the noise-averaged dynamics, enforcing the approximate conservation law of the target Hamiltonian.

	In Fig.~\ref{fig:Noisy}, we compare the ensemble-averaged simulation results for dissipation-free ($\gamma=0$), week ($\gamma=0.2$) and strong  dissipation ($\gamma=0.5$). For $\gamma=0$, ADA-Trotter (orange dots) reaches the better accuracy than Trotter (green triangles), given the same total number of 15 steps and total reachable simulation time $t\approx 5$. With weak dissipation ($\gamma=0.2$), ADA-Trotter still outperforms Trotter but the total reachable time slightly reduces to $t\approx 4.5$. For strong  dissipation ($\gamma=0.5$), ADA-Trotter seems to freeze even at early times, which makes the adaptive methods less efficient. It happens because dissipation generally speeds up the system's heating process towards infinite-temperature and no suitable step size can be identified to self-correct errors in energy and variance once they occur in previous steps. Thorough analysis of dissipation effects will be presented in future works.

\subsection{Dynamics with different initial states for the Ising model}
\label{sec:initial_state}
\begin{figure}[h]
	\centering
	\includegraphics[width=0.7\linewidth]{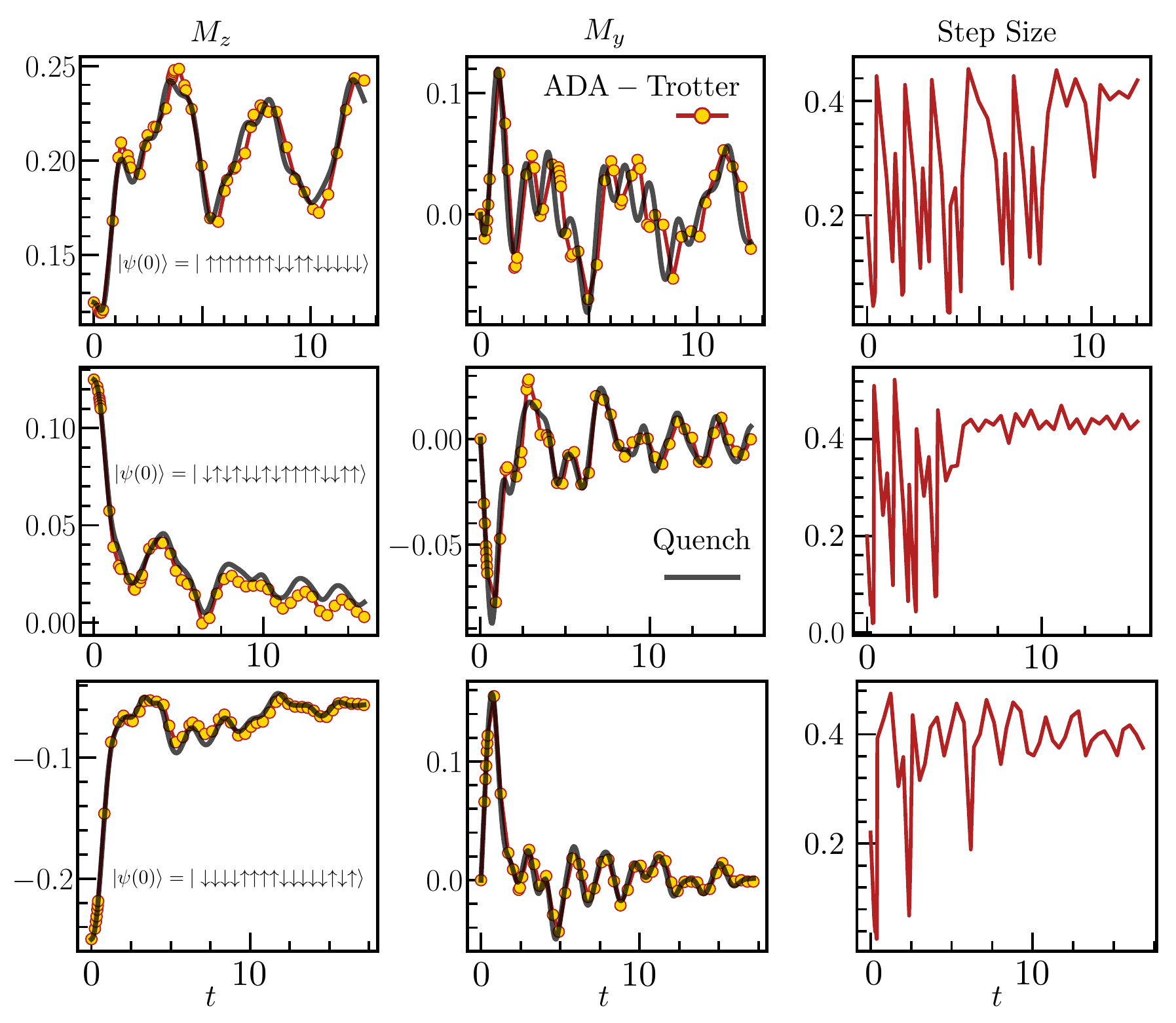}
	\caption{\textbf{[Quantum Ising model]} Dynamics for different initial states which break translation invariance. ADA-Trotter (orange circle) correctly reproduces the exact quench dynamics (black). Step sizes fluctuate in time in the range $[0.01,0.25]$. We use $J_z=1,h_x=0.6,h_z=0.8,L=16,d_{\mathcal{E}}=0.01,d_{\delta\mathcal{E}^2}=0.01$ for numerical simulation. }
	\label{fig:my_label}
\end{figure}

In the main text, the simulation is performed for translation invariant initial states. Here we provide additional results to show that ADA-Trotter also works for different initial states which break the spatial translation symmetry.  In Fig.~\ref{fig:my_label}, we depict the time evolution of magnetization with the same tight controls in both energy and variance. For all simulations the ADA-Trotter uses the bisection method to determine the step sizes and mimics the exact results with weak errors.

\subsection{Long-range Ising model}

\label{sec.long_range}
Some quantum platforms have access to long-range interactions, for instance trapped ions or Rydberg
systems. Here we present numerical simulations of a long-range Ising model; we demonstrate that ADA-Trotter can also be used on these platforms. The Hamiltonian we investigate is
\begin{equation}
	H=\sum_{i<j}^{L} \frac{J_z}{r_{i j}^{\alpha}} \sigma_{i}^{z} \sigma_{j}^{z}+h_x \sum_{i=1}^{L} \sigma_{i}^{x}+h_z \sum_{i=1}^{L} \sigma_{i}^{z},
\end{equation}
with a power-law decaying Ising interaction strength.
The distance between two sites is chosen as $r_{i j}=\min (|i-j|, L-|i-j|)$; hence, periodic boundary conditions can be implemented. We use the same second order Trotter decomposition as in the main text, but now with
\begin{equation}
	H_+=\sum_{i<j}^{L} \frac{J_z}{r_{i j}^{\alpha}} \sigma_{i}^{z} \sigma_{j}^{z}+h_z \sum_{i=1}^{L} \sigma_{i}^{z}, \qquad\qquad    H_-=h_x \sum_{i=1}^{L} \sigma_{i}^{x}.
\end{equation}

\begin{figure}[t!]
	\centering
	\includegraphics[width=0.7\linewidth]{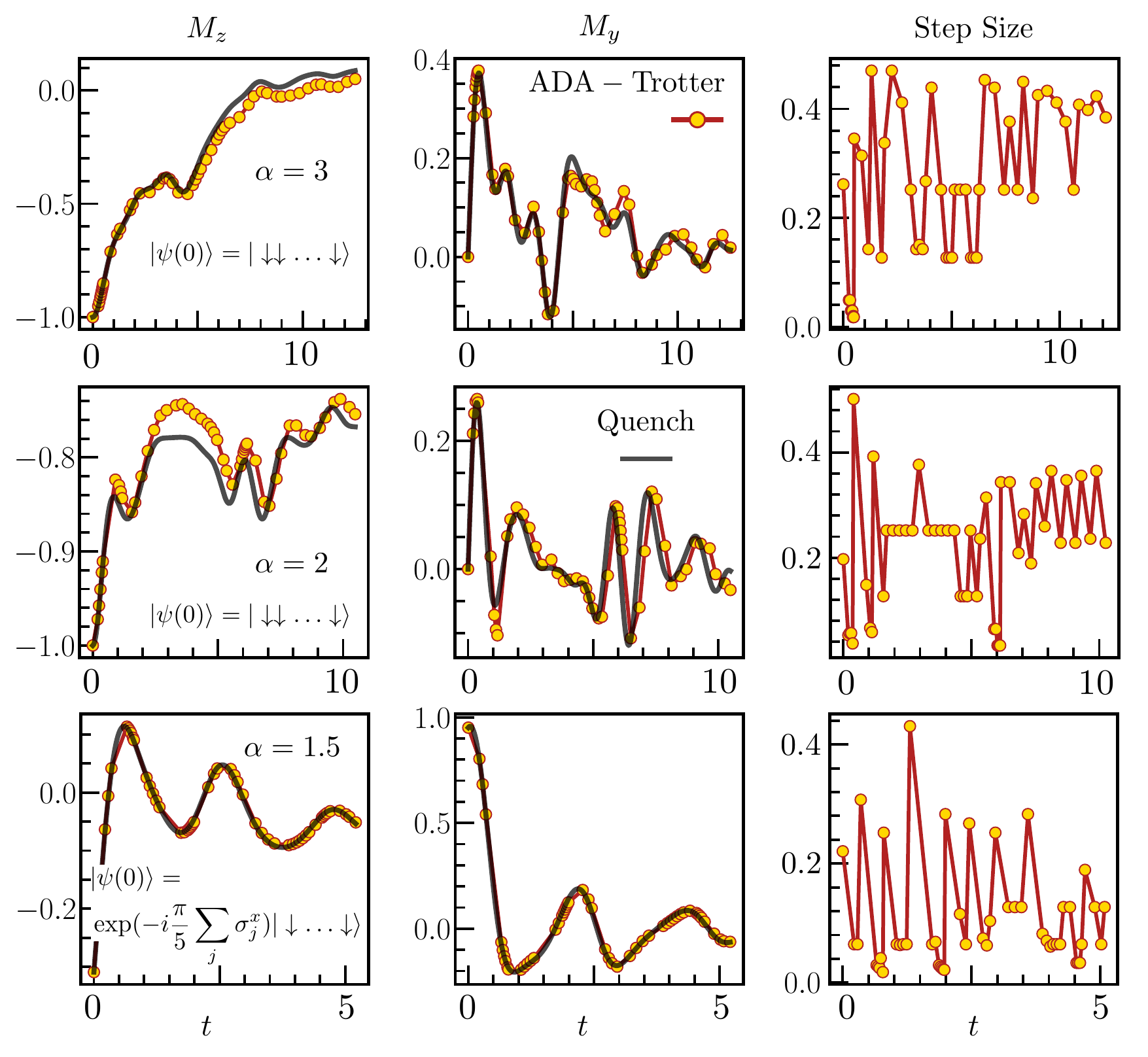}
	\caption{\textbf{[Quantum Ising model with long-range Ising couplings]} Dynamics of local observables and step size. The power law exponent $\alpha$ of the long range interaction is chosen to be 3, 2 and 1.5 from top to bottom rows and ADA-Trotter captures the exact results with weak errors. We use $J_z=1,h_x=0.6,h_z=0.8,L=22$ and $d_{\mathcal{E}}=0.01,d_{\delta\mathcal{E}^2}=0.02$ for numerical simulation.}
	\label{fig:Compare_Flo_Longrange}
\end{figure}

In Fig.~\ref{fig:Compare_Flo_Longrange} we compare the time evolution obtained using ADA-Trotter, with exact quenched results.  We also use tight constraints in energy and variance $d_{\mathcal{E}}=0.01,d_{\delta\mathcal{E}^2}=0.02$, which enable accurate Trotterized time evolution of local observables even when the system has long-range interactions. It is worth noting that long-range interacting systems may not obey ETH. If ETH is violated, the long-time stability of ADA-Trotter is not guaranteed and calls for future investigation.

\subsection{Protection of gauge symmetry in different symmetry sectors for spin-$1/2$ gauge d.o.f.}
\label{sec:gaugesymmetry}
\begin{figure}[h]
	\centering
	\includegraphics[width=\linewidth]{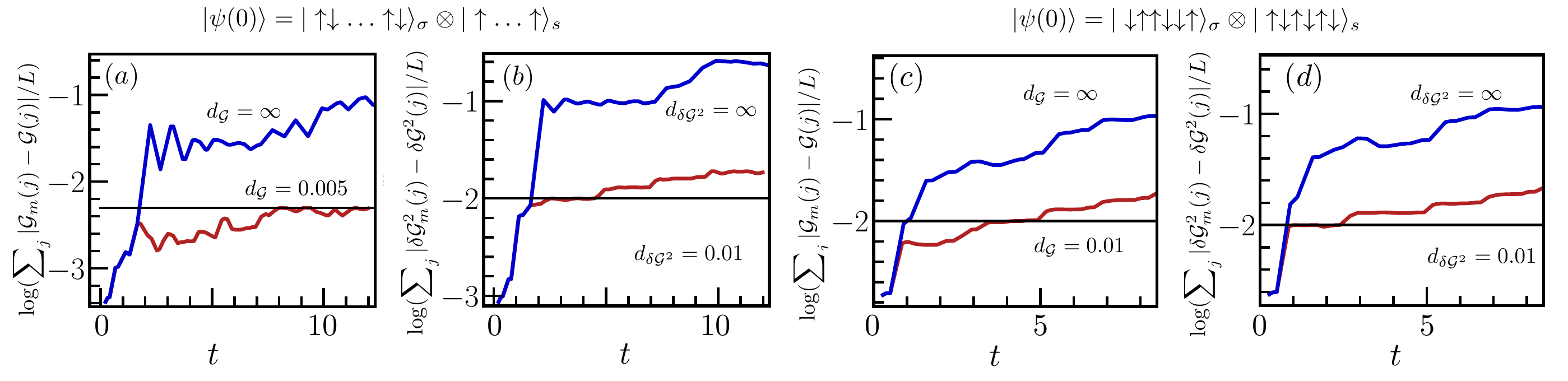}
	\caption{\textbf{[Quantum link model]} Violation of the local gauge symmetry for spin-1/2 gauge d.o.f. in different symmetry sectors. Errors in expectation value of gauge generator are plotted in panel (a) and (c); deviations in the corresponding variance are plotted in panel (b) and (d). Black lines denotes the soft constraints which are used to avoid freezing. Tight constraints slow down the violation of Gauss's law. We use $J=0.5,\mu=0.5,k=0.5,d_{\mathcal{E}}=0.02,d_{\delta \mathcal{E}^2}=0.05,\lambda=0.3, L=6$
		for numerical simulation.
	}
	\label{LGT_Protection_Compare}
\end{figure}

In the main text, we consider spin-1 gauge d.o.f.~and constraints on the gauge violation are defined for the symmetry sector with $G_j|\psi(0)\rangle=0$ for any $j$. The symmetry protection can indeed be generalized for other spin-$S$ d.o.f.~and states in any symmetry sector with non-zero initial expectation value of $G_j$. We define the general constraints on symmetry violation as 
\begin{equation}
	\sum_{j}|\mathcal{G}_{m}(j)-\mathcal{G}(j)|/L < d_\mathcal{G}, \qquad\qquad 
	\sum_{j} |\delta \mathcal{G}^2_{m}(j)-\delta \mathcal{G}^2(j)|/L < d_{\delta \mathcal{G}^2} \; .
	\label{eq:e_constraint_gauss}
\end{equation} 
Here we define the expectation value of the gauge symmetry generator and its variance as
$\mathcal{G}(j)=\langle \psi(0)| G_j|\psi(0) \rangle$, $\delta \mathcal{G}^2(j)=\langle \psi(0) | G_j^2 |\psi(0) \rangle - \mathcal{G}^2(j)$ for the initial state $|\psi(0) \rangle$, and  $\mathcal{G}_m(j)=\langle G_j\rangle_{m}$, $\delta \mathcal{G}_m^2(j)=\langle G_j^2 \rangle_{m} - \mathcal{G}_m^2(j)$ for the state $|\psi (t_m) \rangle$ obtained by the ADA-Trotter at time $t_m$. As shown in Fig.~\ref{LGT_Protection_Compare}, for different initial states,
the tight constraints with small $d_{\mathcal{G},\delta \mathcal{G}^2}$ significantly slow down the gauge violation. 

As mentioned in the main text, a tight constraint in conserved quantities can cause ADA-Trotter to `freeze', namely ADA-Trotter tends to always choose the smallest possible step size allowed in the searching algorithm, see details in Sec.~\ref{app:Search_Algorithm}. Consequently, the quantum state barely propagates, despite the large number of quantum gates consumed. To resolve this issue, rather than using fixed tolerances, here we employ a soft constraint such that tolerances can increase by $30\%$ of their original values whenever the smallest step size is chosen. 

More concretely, in sequential search (see Sec.~\ref{app:Search_Algorithm}), we introduce the control function
	\begin{equation}
		\label{eq.condition_SM_gauge}
		f_{\mathcal{G}}(\delta t_m) = \sum_{j}|\mathcal{G}_{m+1}(j)-\mathcal{G}(j)|/L - d_\mathcal{G},  \ f_{\delta \mathcal{G}^2}(\delta t_m) = \sum_{j} |\delta \mathcal{G}^2_{m+1}(j)-\delta \mathcal{G}^2(j)|/L - d_{\delta \mathcal{G}^2} .
	\end{equation}
	in addition to Eq.~\ref{eq.condition_SM} to track the error in Gauss's law. If the one of the four (2 for energy and 2 for Gauss's law) constraints are too tight, say $f_{\mathcal{G}}(\delta t_m)>0$ even for the smallest step size $t_{\mathrm{min}}$, we will choose $t_{\mathrm{min}}$ to propagate the state to time $t_m+t_{\mathrm{min}}$, and update the constraint as $d_\mathcal{G}\to 1.3 d_\mathcal{G}$ which will be used for all future searching process. Constraints on other conservation law remain unchanged.

\section{Technical details for ADA-Trotter}
\subsection{Search Algorithm\label{app:Search_Algorithm}}
There are different search algorithms to propose a smaller trotter step size for ADA-Trotter at each time, in order to satisfy the constraints in conserved quantities. Here we illustrate two possibilities, namely the sequential search, which identifies the largest possible step size in a given temporal range, and bisection search, which reduces the number of  attempts needed for the searching process but may not identify the globally optimal step size. On real quantum simulators, the measurement process of conserved quantities can also be costly and increase the runtime. An efficient search algorithm, for instance the bisection method, can be particularly useful for practical purposes.

For a given quantum state $\ket{\psi(t_m)}$, we aim to find a large time step $\delta t_m$ to satisfy
$
|\mathcal{E}_{m+1}-\mathcal{E}| < d_\mathcal{E}, \ |\delta \mathcal{E}^2_{m+1}-\delta \mathcal{E}^2| < d_{\delta \mathcal{E}^2},
$
where $\mathcal{E},\delta\mathcal{E}^2$ correspond to energy and variance density for the initial state.
These conditions ensure the conservation of the average density and its fluctuations up to the maximally allowed errors $d_\mathcal{E}$ and $d_{\delta \mathcal{E}^2}$.
To achieve these conditions at each time $t_m$, we define the function
\begin{equation}
	\label{eq.condition_SM}
	f_{\mathcal{E}}(\delta t_m) = |\mathcal{E}_{m+1}-\mathcal{E}| - d_\mathcal{E},  \ f_{\delta \mathcal{E}^2}(\delta t_m) = |\delta \mathcal{E}^2_{m+1}-\delta \mathcal{E}^2| - d_{\delta \mathcal{E}^2}.
\end{equation}
At short times, before the energy and variance saturate at their tolerances, for a positive but small error bound $d_{\mathcal{E}},d_{\delta \mathcal{E}^2}$, we know $f_{\mathcal{E}}( \delta t_m)\to -d_{\mathcal{E}} < 0$ and  $f_{\delta \mathcal{E}^2}( \delta t_m)\to -d_{\delta \mathcal{E}^2}<0$ in the limit $\delta t_m\to 0$. Also, for large $\delta t_m$, Trotterized unitary $U_T(\delta t_m)$ can be completely different from the target unitary and cause large errors, hence $f_{\mathcal{E},\delta \mathcal{E}^2}( \delta t_m)>0$. There exists a solution $\delta t_m$ such that  $f_{\mathcal{E},\delta \mathcal{E}^2}( \delta t_m)=0$ for which one can search by the following sequential search algorithm.

\begin{algorithm}[H]
	\caption{Sequential search}
	\label{alg:sequential}
	\begin{algorithmic}
		
		\State {\textbf{Input} Function $f_{\mathcal{E},\delta \mathcal{E}^2}( \delta t_m)$ at a given time $t_m$, 
			endpoint for the time window $t_{\mathrm{min}}$ and $t_{\mathrm{max}}$, resolution of time $\delta \tau$, total number of attempts is $N_{\mathrm{max}}=(t_{\mathrm{max}}-t_{\mathrm{min}})/\delta \tau$, }
		
		\State {\textbf{Conditions} $t_{\mathrm{min}} < t_{\mathrm{max}},f_{\mathcal{E},\delta \mathcal{E}^2}( t_{\mathrm{min}})<0,f_{\mathcal{E},\delta \mathcal{E}^2}( t_{\mathrm{max}})>0$.}
		
		\State {\textbf{Output} Largest value of $\delta t_m$ which satisfies $f_{\mathcal{E},\delta \mathcal{E}^2}(\delta t_m)<0$.}
		\\
		\State $N \gets 1,\delta t_m\gets t_{\mathrm{max}}$
		
		\While{$N \leq N_{\mathrm{max}}$} \Comment{Limit iterations to prevent infinite loop and negative values of $\delta t_m$}
		
		\If {$f_{ \mathcal{E}}(\delta t_m)<0$ \textbf{and} $f_{\delta \mathcal{E}^2}(\delta t_m)<0$}
		
		\State {\textbf{stop} \Comment{Stop search algorithm if both conditions are satisfied.}}
		\Else 
		\State {$\delta t_{m}\gets \delta t_m-\delta \tau$ \Comment{Reduce the trotter step size.}}
		\EndIf 
		\State $N\gets N+1$
		\EndWhile  
	\end{algorithmic}
\end{algorithm}

The sequential search algorithm shown in Algorithm~\ref{alg:sequential} identifies the largest possible trotter step size for which both constraints in energy and variance are satisfied, up to the resolution in time $\delta \tau$. However, improving the resolution by using a smaller $\delta \tau$ can be costly as normally the attempt times increases linearly with $N_{\mathrm{max}}$. To improve the efficiency, bisection search can be considered. To find the root of a single continuous and monotonic function, one gets the attempt times scaling $\log(N_{\mathrm{max}})$. Here, we generalize it to find a solution satisfying two conditions, by performing conventional bisection search in energy or variance iteratively as detailed in Algorithm~\ref{alg:bisection}. Both algorithms can be used to select suitable step sizes adaptively in time. As shown in Fig.~\ref{fig:attempt_compare1} (a) and (c) with different energy and variance tolerances, both methods generate reliable local time evolution with weak errors.  

\begin{algorithm}[H]
	\caption{Bisection search}\label{alg:bisection}
	\begin{algorithmic}
		\State \textbf{Input} Function $f_{\mathcal{E},\delta \mathcal{E}^2}( \delta t_m)$ at a given time $t_m$, precision of the search algorithm $p_{\mathcal{E},\delta \mathcal{E}^2}$, 
		endpoint for the time window $t_{\mathrm{min}}$ and $t_{\mathrm{max}}$, maximum iterations $M_{\mathrm{max}}$ to search for the solution of either $f_{\mathcal{E}}$ or $f_{\delta \mathcal{E}^2}$, maximum time $R_{\mathrm{max}}$ for cross check.
		\State \textbf{Conditions} $t_{\mathrm{min}} < t_{\mathrm{max}},f_{\mathcal{E},\delta \mathcal{E}^2}( t_{\mathrm{min}})<0,f_{\mathcal{E},\delta \mathcal{E}^2}( t_{\mathrm{max}})>0$.
		\State \textbf{Output} value of $ t_{\mathrm{mid}}$ which differs from a root of $f_{\mathcal{E},\delta \mathcal{E}^2}( t_{\mathrm{mid}})=0$ by less than $p_{\mathcal{E},\delta \mathcal{E}^2}$.
		\\
		
		\State $M \gets 1,R \gets 1$, $t_{\mathrm{mid}} \gets t_{\mathrm{max}}$
		\While{$R \leq R_{\mathrm{max}}$} \Comment{Limit iterations to prevent infinite loop for cross checks}
		\If {$R$ is even}  \Comment{Switch between energy and variance  examination}
		\State $f=f_{\mathcal{E}},p=p_{\mathcal{E}}$
		\Else{ $f=f_{\delta \mathcal{E}^2},p=p_{\delta \mathcal{E}^2}$ }
		\EndIf
		
		\If {$f(t_{\mathrm{mid}})<p, R\neq 1$}
		\State {\textbf{break}} \Comment{Break the while loop of $R$ when two conditions are satisfied consecutively.}
		\Else {
			\While{$M \leq M_{\mathrm{max}}$}\Comment{Limit iterations to prevent infinite loop for bisection search}
			\State $t_{\mathrm{mid}}\gets (t_{\mathrm{min}}+t_{\mathrm{max}})/2$
			\If{$|f(t_{\mathrm{mid}})|<p$}
			\State $t_{\mathrm{max}}\gets t_{\mathrm{mid}}$ \Comment{Update the endpoint for the  time window to exam the other condition.}
			\State \textbf{break} \Comment{Break the while loop of $M$ when one condition is satisfied.}
			\EndIf 
			\If{$\mathrm{sign}(f(t_{\mathrm{mid}})) = \mathrm{sign}(f(t_{\mathrm{min}}))$}
			\State {$t_{\mathrm{min}}\gets t_{\mathrm{mid}}$}
			\Else {\ $t_{\mathrm{max}}\gets t_{\mathrm{mid}}$} \Comment{Update the searching interval by halving.}
			\EndIf 
			\State $M\gets M+1$
			\EndWhile  
		}
		\EndIf 
		\State $R\gets R+1$
		\EndWhile 
	\end{algorithmic}
\end{algorithm}

\subsection{Comparation between different searching algorithms\label{app:Search_Algorithm_compare}}
\begin{figure}[h]
	\centering
	\includegraphics[width=0.9\linewidth]{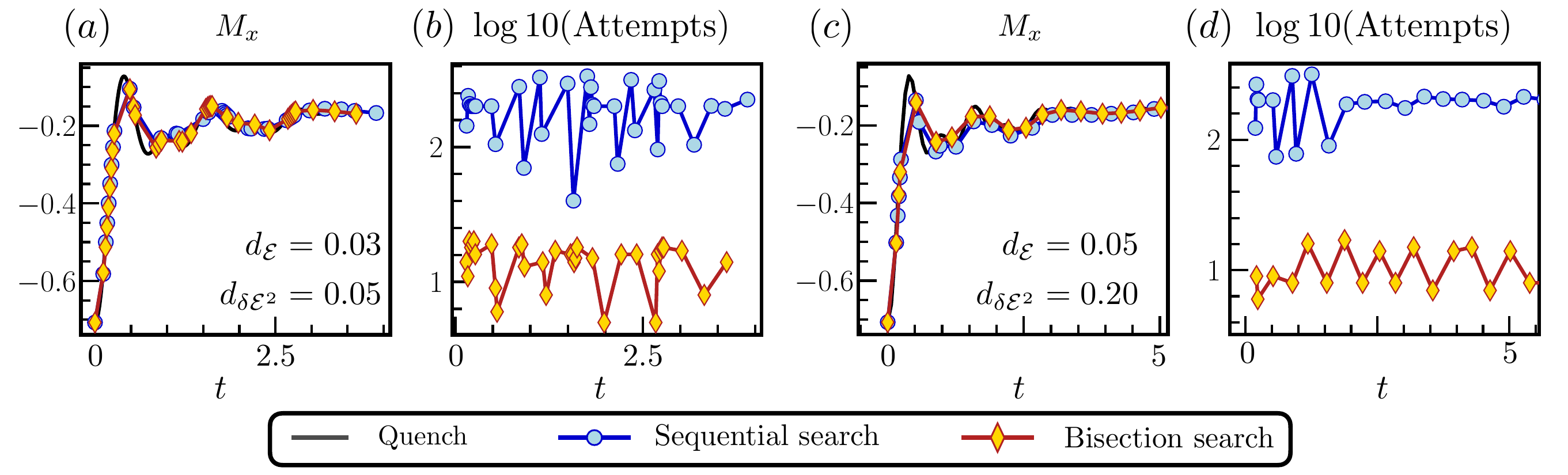}
	\caption{(a,c) Local time evolution. (b,d) Number of attempts in the search process. Bisection search uses fewer attempts than sequential search. We use $J_z=-1,h_x=-1.7,h_z=0.5,L=20$ and initial state $\exp(-i\pi/8\sum_j\sigma_j^y)\bigotimes_i\ket{\downarrow}_i$ for numerical simulation. Precision for the bisection method is $p_{\mathcal{E},\delta \mathcal{E}^2}=d_{\mathcal{E},\delta \mathcal{E}^2}/30$. Resolution of time $\delta \tau$ for sequential search is 0.001.}
	\label{fig:attempt_compare1}
\end{figure}

Now we numerically analyse the attempt times with different searching algorithms. In Fig.~\ref{fig:attempt_compare1} (b) and (d) where different constraints are used, we plot the number of attempts for both sequential(blue) and bisection(orange) search versus time. In both panels, if a high resolution in time is used ($\delta \tau=0.001$) sequential search can be expensive as it uses more than 100 attempts (a log scale is used in the plot) to identify the suitable step. In contrast, bisection search is very efficient where attempt number drops to approximately 15 on average, one order of magnitude smaller than sequential search. Importantly, a direct consequence is the reduction of the measurement times required after each attempt to examine whether errors in energy and variance are bounded. In addition, for sequential search, both measurements of energy and its variance are required, whereas for bisection search only one of them is needed in each $R$ loop, which saves experimental efforts remarkably.

\begin{figure}[h]
	\centering
	\includegraphics[width=0.95\linewidth]{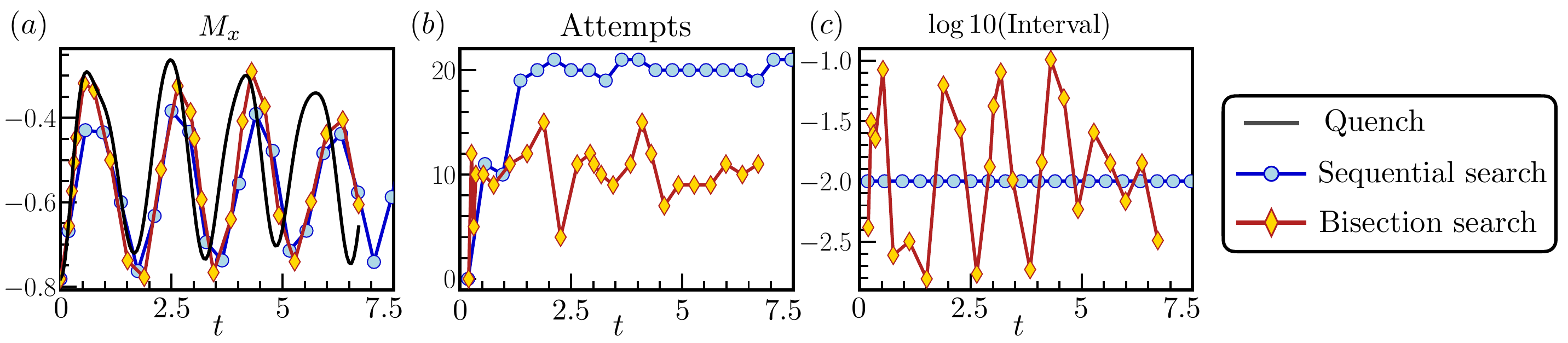}
	\caption{
		(a) Local time evolution. (b) Number of attempts in the search process. (c) Resolution in time for different search algorithm. Bisection method achieves the same resolution as sequential scheme, but fewer attempts to search for the step size. We use $J_z=-1,h_x=1.2,h_z=0.6,L=20$ and initial state  $\exp(-i\pi/7\sum_j\sigma_j^y)\bigotimes_i\ket{\downarrow}_i$ for numerical simulation. Precision for the bisection method is $p_{\mathcal{E},\delta \mathcal{E}^2}=d_{\mathcal{E},\delta \mathcal{E}^2}/10$ with $d_{\mathcal{E}}=0.05,d_{\delta \mathcal{E}^2}=0.1.$ Resolution of time $\delta \tau$ for sequential search is 0.01.}
	\label{fig:attempt_compare2}
\end{figure}

Number of attempts for sequential search can also be reduced by using a larger value $\delta \tau$. In Fig.~\ref{fig:attempt_compare2} we choose $\delta \tau=0.01$ and the resultant attempts also drops by one order of magnitude compared with Fig.~\ref{fig:attempt_compare1}. One can also define the resolution in time for the bisection search by using $(t_{\mathrm{min}}-t_{\mathrm{max}})/2$ at the end of the search, as plotted in panel (c) in orange. It fluctuates in time around the average value $\sim10^{-2}$ which is roughly the same as in sequential search(blue). However, bisection search is still more efficient as it only requires half of the attempts as shown in panel (b).

\subsection{System size dependence of bisection search\label{app:Search_Algorithm_size}}
\begin{figure}[h]
	\centering
	\includegraphics[width=0.95\linewidth]{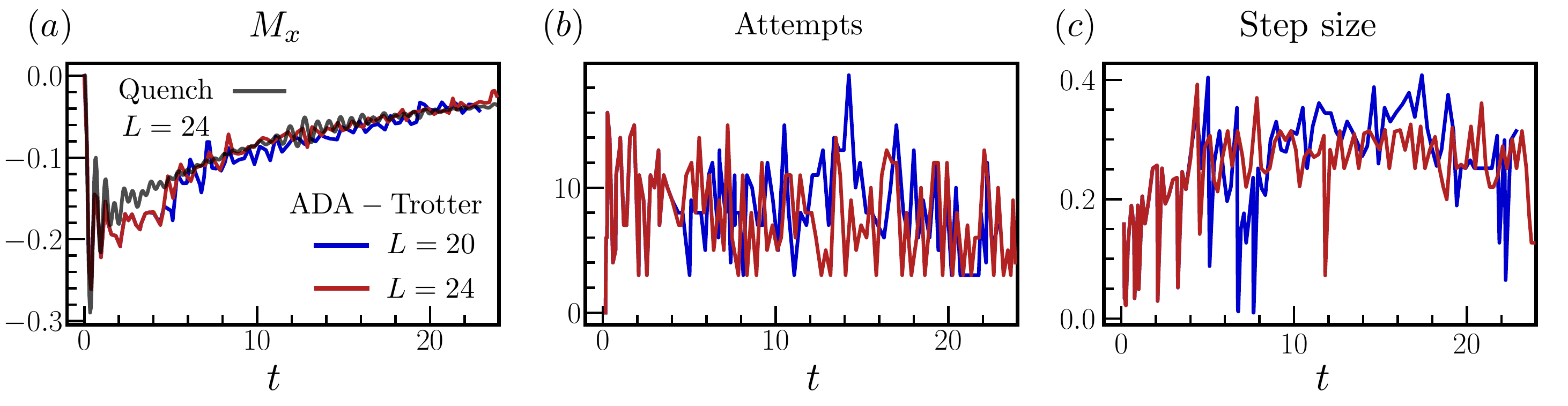}
	\caption{(a) Local time evolution. Results converge for larger system sizes. (b) Number of attempts in the search process by bisection method.  (c) Selected step size. Attempt number does not scale up for larger system sizes, suggesting that ADA-Trotter can also be implemented efficiently on quantum simulators with a large number of qubits. 
		We use $J_z=-1,h_x=-2,h_z=0.2,L=24$ and initial state polarized in negative $y$ direction for the numerical simulation. Precision for the bisection method is $p_{\mathcal{E},\delta \mathcal{E}^2}=d_{\mathcal{E},\delta \mathcal{E}^2}/10$ with $d_{\mathcal{E}}=0.05,d_{\delta \mathcal{E}^2}=0.1.$}
	\label{fig:finitesize}
\end{figure}

We notice that the number of attempts for bisection search does scale up for larger system sizes. Therefore, this search algorithm can also be efficiently implemented on quantum simulators, which ideally have a much larger number of accessible qubits. 
	We illustrate this property in Fig.~\ref{fig:finitesize} (b) where results for different system sizes $L=20$ (blue) and $24$ (red) are presented. At short times ($t\le5$) dynamics only occurs locally, hence, attempt numbers (b) are insensitive to system size. In contrast, at later times, attempt number are generally different for different system sizes, but it always fluctuates around a constant number and does not scale up for larger $L$. The selected step sizes are generally distinct for different system sizes as shown in Fig.~\ref{fig:finitesize} (c) and do not seem to converge for larger $L$. It happens because at each time ADA-Trotter needs to adjust the step size differently to correct errors in mean energy and variance, which would rely on the system sizes. Although step sizes may be different, the resulting local time evolution already converges to the desired solution (black in Fig.~\ref{fig:finitesize} (a)). 
\subsection{Time window in bisection search\label{app:Search_Algorithm_window}}
\begin{figure}[h]
	\centering
	\includegraphics[width=0.9\linewidth]{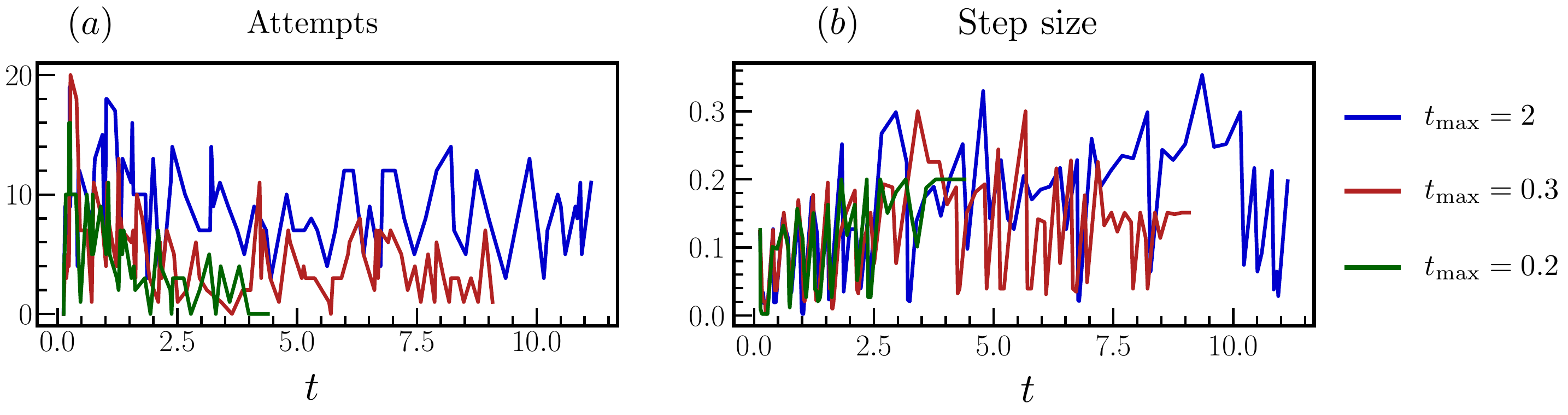}
	\caption{(a) Number of attempts in the search process by bisection method.  (b) Selected step size. Attempt number reduces for a smaller searching time window, but it also limits the total achievable simulation time. 
		We use $J_z=-1,h_x=-2,h_z=0.2,L=20$ and initial state polarized in negative $y$ direction for the numerical simulation. Precision for the bisection method is $p_{\mathcal{E},\delta \mathcal{E}^2}=d_{\mathcal{E},\delta \mathcal{E}^2}/10$ with $d_{\mathcal{E}}=0.02,d_{\delta \mathcal{E}^2}=0.05.$}
	\label{fig:timewindow}
\end{figure}
Number of attempts also depends on the time window $[t_{\mathrm{min}},t_{\mathrm{max}}]$ in the bisection algorithm. In our simulation, $t_{\mathrm{min}}$ is normally fixed to be a positive small number, e.g., $t_{\mathrm{min}}=0.01$, such that the system can still propagates by a small but nonzero time step even when the algorithm freezes.  For the upper bound $t_{\mathrm{max}}$ of the time window, intuitively, a larger $t_{\mathrm{max}}$ encourages ADA-Trotter to find the largest possible step size without violating the energy constraints. However, such a strategy generally increases the number of attempts during search and the measurement cost (see details in the next section), hence making ADA-Trotter less efficieint. As shown in Fig.~\ref{fig:timewindow} (a), if we only allow a small time window for searching, e.g., $t_{\mathrm{max}}=0.2$ (green), very few number of attempts are needed, especially for $t\ge 2$ where approximately 2 attempts on average are made. However, it significantly limits the total simulation time that ADA-Trotter can achieve with the fixed number of accessible gates. Therefore, we expect that there may exist certain optimal searching time window that can achieve sufficiently long simulation time while the number of attempts are experimentally feasible. However, such an optimal $t_{\mathrm{max}}$ should highly depend on the models, initial states and it may also be adaptive in time. 

\subsection{Measurement cost of ADA-Trotter}
\label{sec.cost}
Here we discuss the extra measurement cost of making the step size adaptive. At each trotter step, one needs to measure the conserved quantities, and search for the suitable step size based on the measurement outcomes. The measurement cost highly depends on the DQS platforms and the target models. For instance, Ref.~\cite{kokail2019self} discussed the measurement process for a lattice Schwinger model on a trapped-ion quantum simulator. There,  simultaneous projective measurements of all qubits can be preformed in the logical z-basis, via spatially
resolved fluorescence. Single qubit rotation is used to enable projective measurement in different product bases. To obtain energy, only a constant number of measurement bases are
needed; for the variance, the required bases number scales only linearly in system size, which is quite efficient. Recently, more efficient measurement schemes based on randomized measurements and classical shadows have been proposed~\cite{elben2023randomized}. The measurement cost for the Hamiltonian variance has been improved from linear to logarithmic in system size~\cite{huang2020predicting}. Additionally, by derandomizing the randomized protocol, further significant improvements can be achieved`\cite{huang2021efficient}. Also, one may employ Bayesian optimization to estimate the expectation values based on few number of measurement shots~[79] to minimize the required experimental efforts. Concrete measurement protocols on different digital devices would be worthwhile to explore in the future for practical usage.

\subsection{Full dynamics for Fig.~\ref{fig:schematic}}
\label{sec:full_time}
\begin{figure}[h]
	\centering		\includegraphics[width=0.9\linewidth]{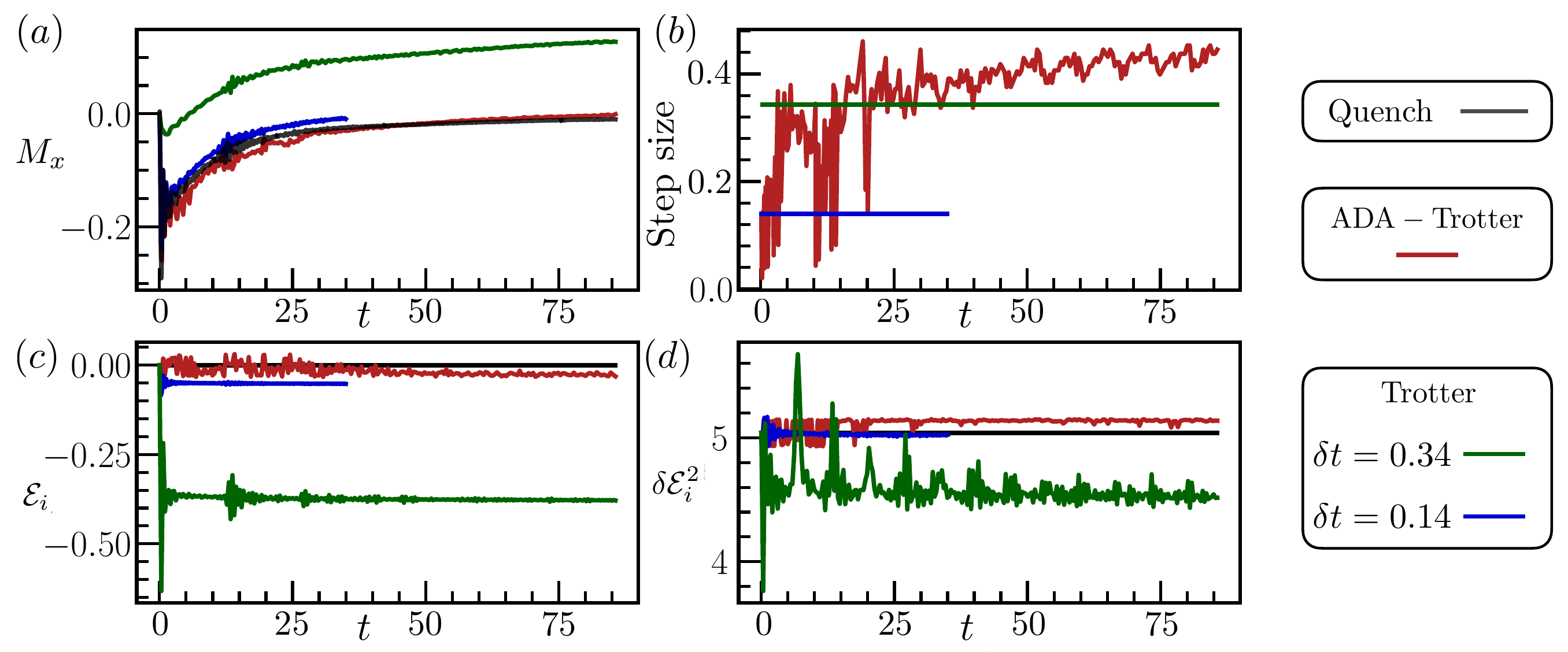}
	\caption{(a) Magnetization in $x$ direction evolved by different trotter schemes. Total number of Trotter steps is fixed to be 250. ADA-Trotter reproduces the local time evolution throughout all simulation times. However, to achieve the same simulation time, conventional methods (green, $\delta t=0.34$) exhibits notable errors even at very early times. A small step size is used ($\delta t=0.14$) which suppresses the local errors, but the total achievable time is limited. (b) Trotter step size $\delta t_m$. A fixed step size is chosen for fixed step Trotter whereas adaptive $\delta t_m$ is used for ADA-Trotter.  Expectation value and variance of the target Hamiltonian are plotted in panel (c) and (d). ADA-Trotter conserves the Hamiltonian better than conventional Trotter. We use $J_z=-1,h_x=-2,h_z=0.2,d_{\mathcal{E}}=0.03,d_{\delta\mathcal{E}^2}=0.1,L=24$ and initial state polarized in negative y direction for numerical simulation. Bisection search is used and we choose $p_{\mathcal{E},\delta \mathcal{E}^2}=d_{\mathcal{E},\delta \mathcal{E}^2}/10$.}
	\label{fig:Compare_Flo_examples1}
\end{figure}
\begin{figure}[h]
	\centering		\includegraphics[width=0.65\linewidth]{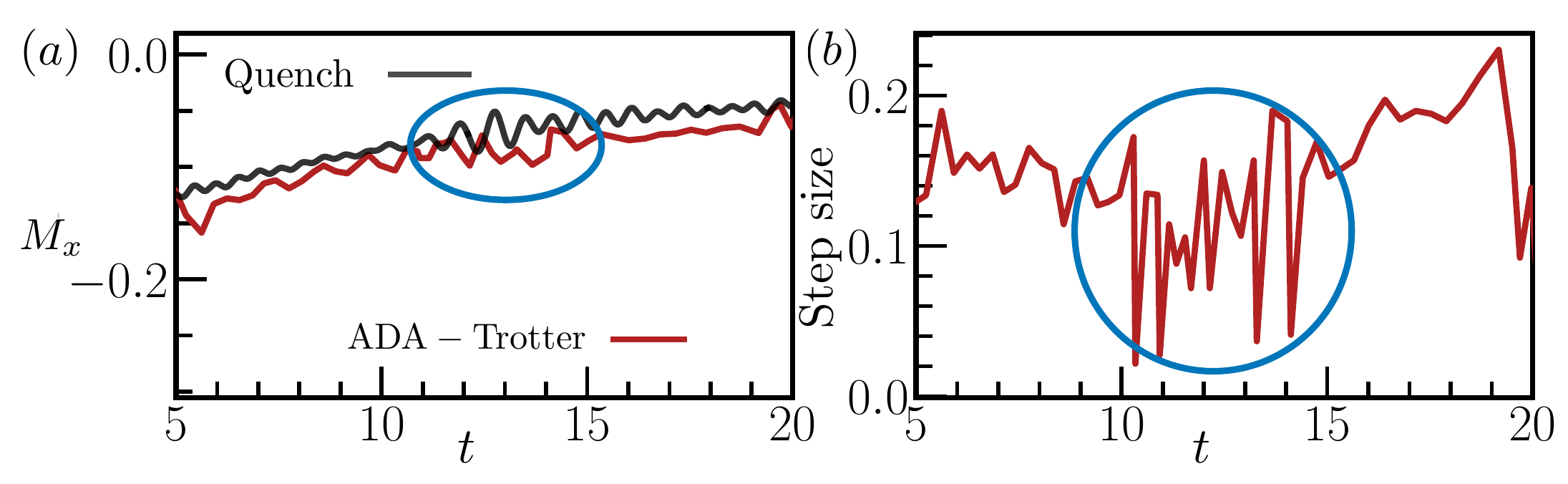}
	\caption{(a) Magnetization in $x$ direction. (b) Trotter step size $\delta t_m$ fluctuates by one order of magnitude to capture the quick oscillations in the local observable. We use the same parameters as in Fig.~\ref{fig:Compare_Flo_examples1}}
	\label{fig:Compare_Zoomin}
\end{figure}

Here we plot the full time evolution for Fig.~\ref{fig:schematic} in Fig.~\ref{fig:Compare_Flo_examples1} to show that ADA-Trotter(red) captures the exact local time evolution (black) with weak errors for all simulation times. Energy are plotted in panel (c) where adaptive step size leads to time-dependent energy dynamics which fluctuate around the correct value, also the similar behavior occurs in panel (d) for the energy variance. It suggests that ADA-Trotter can correct simulation errors by modulating step sizes, hence preserving the target Hamiltonian better than the fixed step Trotter(green) if the same simulation time is achieved. It is worth noting that the selected step sizes can exhibit highly non-monotonic behavior. In Fig.~\ref{fig:Compare_Zoomin}, we use the same data as in Fig.~\ref{fig:Compare_Flo_examples1} but concentrate on the time window between $t=10$ to 15. As illustrated in panel (a), the system exhibits relatively stronger oscillations in local observable in this time window, and ADA-Trotter reduces the step sizes by approximately one order of magnitude to properly capture this dynamical feature. 

\subsection{Verification of variance condition}
\label{sec.variance_condition}
\begin{figure}[h]
	\centering
	\includegraphics[width=0.6\linewidth]{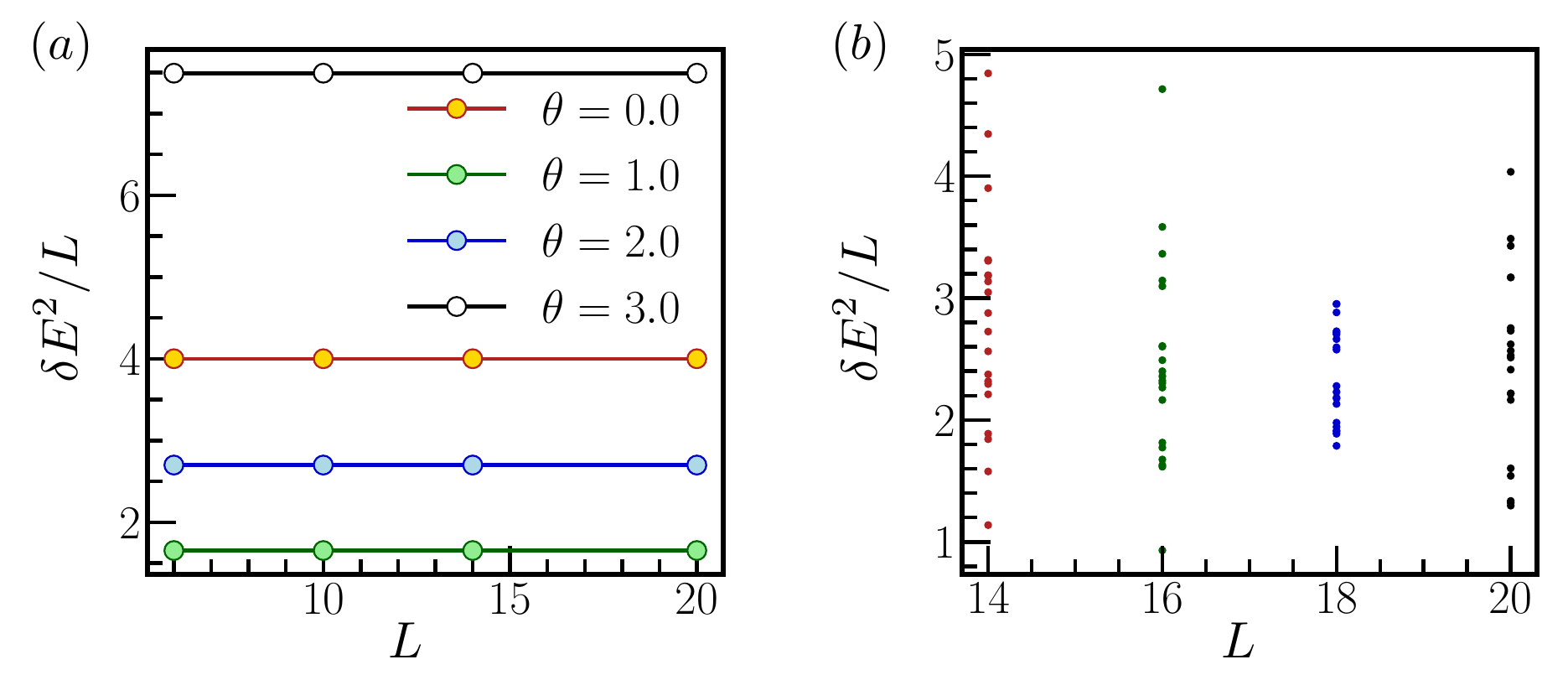}
	\caption{Verification of the condition $\delta E^2\geq aL$ with a positive $a$, for translation invariant states in panel (a) and random states in panel (b). We use the Hamiltonian parameters $J_{zz}=-1,h_x=2,h_z=1$ in (a) and $J_{zz}=-1,h_x=-1,h_z=1.6$ in (b).}
	\label{fig:variancebound}
\end{figure}
It has been proved that for a product state $\ket{\psi(0)}$ with mean energy $E_0=$ and energy variance $\delta E^2$, such that the variance is lower bounded by the number of qubits $\delta E^2\ge aL$ with $a>0$, the energy distribution converges to a Gaussian $\rho(E)=\frac{1}{\sqrt{2 \pi} \delta E} e^{-\left(E-E_0\right)^2 / 2 \delta E^2}$ in the thermodynamic limit~\cite{hartmann2004gaussian}. Here we numerically verify the condition $\delta E^2\ge aL$ with $a>0$ with respect to the Hamiltonian
	$
	H= J_z \sum_{j} \sigma_{j}^{z} \sigma_{j+1}^{z}+h_z \sum_{j} \sigma_{j}^{z}+h_x \sum_{j} \sigma_{j}^{x},
	$
	for two cases: (a) translation invariant product states $\ket{\psi(0)}=\exp(-i\sum_j\theta)|\downarrow\dots\downarrow\rangle$ with periodic boundary condition; (b) spatially random states $\ket{\psi(0)}=\exp(-i\sum_j\theta_j)|\downarrow\dots\downarrow\rangle$ with open boundary condition, where $\theta_j$ is randomly chosen from $[0,\pi]$. As shown in Fig.~\ref{fig:variancebound}, both panels suggest that $\delta E^2/L$ is lower bounded by a positive constant. Therefore, we expect that the energy distribution of those states will converge to Gaussian distribution for $L\to\infty$.

\subsection{Dynamics of higher moments}
\label{sec:higher_moments}
\begin{figure}[h]
	\centering
	\includegraphics[width=\linewidth]{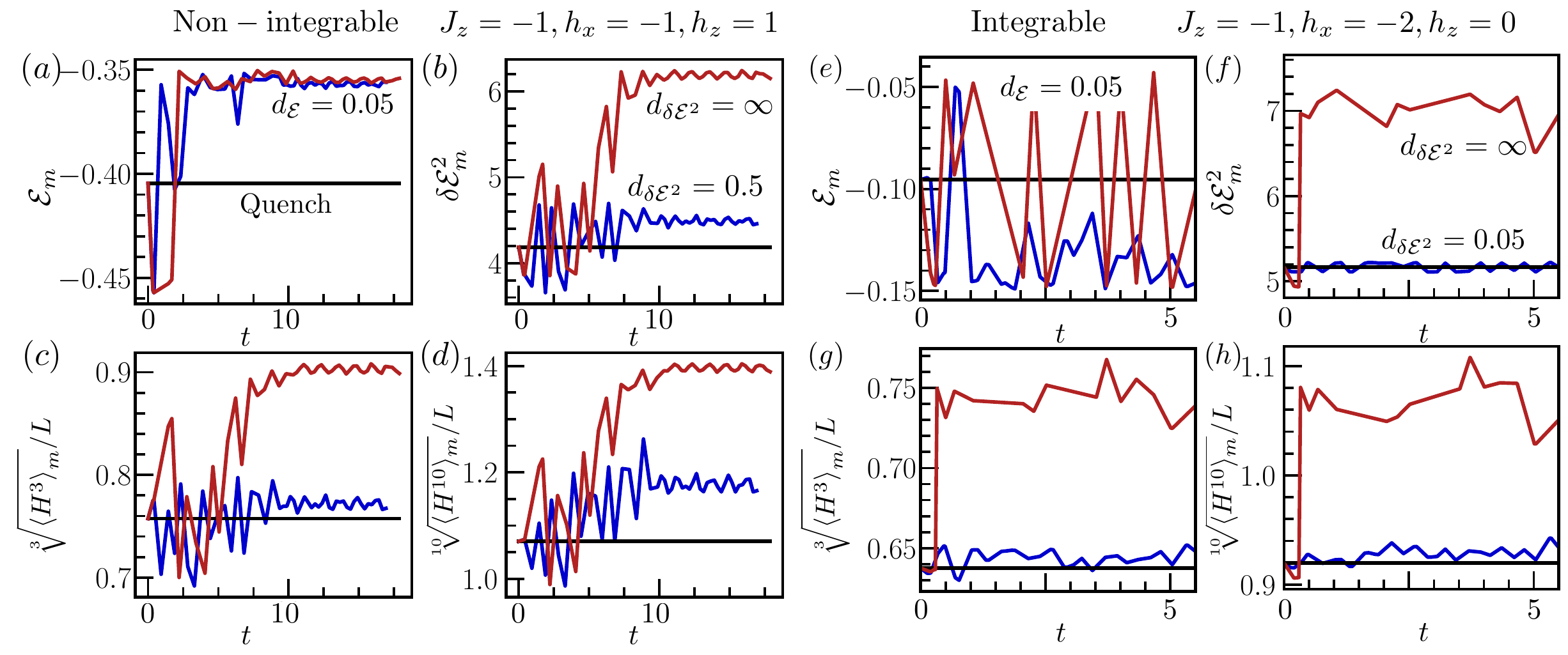}
	\caption{Dynamics of expectation values of different moments of the Hamiltonian. Constraining the mean and variance of the energy (blue) is sufficient to bound the deviation of higher moments. Only constraining energy (red) is not enough.  Both integrable (right) and non-integrable (left) systems are considered. We use $L=22$   for numerical simulation.}
	\label{fig:powersH}
\end{figure}
In this section we give more examples showing that constraining the violation in energy and variance suffices to preserve higher moments with ADA-Trotter. In Fig.~\ref{fig:powersH}, we consider the initial state $\exp(i\pi\sum_j\sigma_j^y/5){|\downarrow\dots \downarrow\rangle}$ and simulate non-integrable Hamiltonian in panels (a-d), and an integrable system in panels (e-h). In both cases, higher moments are better preserved when a tight constraint in the variance is employed. As discussed in Sec.~\ref{sec.variance_condition}, it happens because for a product state, the energy distribution converges to a Guassian distribution for sufficiently large system sizes. 

\begin{figure}[h]
	\centering
	\includegraphics[width=0.9\linewidth]{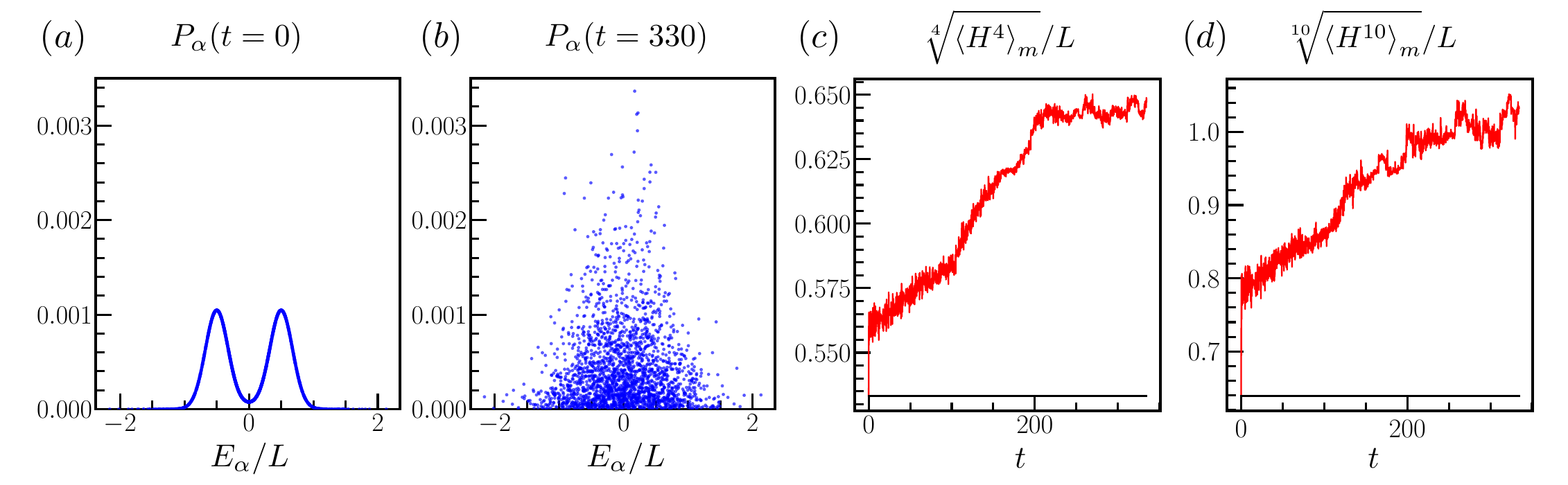}
	\caption{Energy distribution and higher moments of the Hamiltonian for an initial state that is far from a Gaussian distribution. (a) and (b) show the energy distribution at early ($t=0$) and long times ($t=330$). (c) and (d) depict the higher moments where clear deviation from the exact conserved quantity (black) can be observed. Hence, for states that are far from a Gaussian distribution, controlling the lowest two moments of the target Hamiltonian may not be sufficient for ADA-Trotter to realize reliable digital simulation. We use $J_{z}=-1, h_x=-2,h_z=0.2, d_{\mathcal{E}}=0.01,d_{\delta\mathcal{E}^2}=0.1,L=16$ for numerical simulation.}
	\label{fig:powersH_Cat}
\end{figure}
In contrast, in Fig.~\ref{fig:powersH_Cat}, we consider a ``cat state" in the eigenbasis that is far from a Gaussian distribution. More concretely, the initial state is defined as $\ket{\psi(0)} =(1/C)\sum_{\alpha}[g(E_1,E_{\alpha})+g(E_2,E_{\alpha})]\ket{E_{\alpha}}$, where $C$ is the normalization factor and the function $g$ is the Gaussian function $g(x_0,x,w) = \exp (-(x-x_0)^2/(2w^2)) $ . The overall energy distribution will be far from a Gaussian distribution if $E_1$ and $E_2$ have a large difference and the width $w$ is sufficiently small. Numerically, we choose $E_1=0.5L, E_2=-0.5L,w=\sqrt{2L}$, resulting in the energy distribution $P_{\alpha}=|\langle \psi(0)|E_{\alpha}\rangle|^2$ as shown in Fig.~\ref{fig:powersH_Cat}(a). At long times, e.g. $t=330$ in panel (b), the distribution becomes completely different from the initial state, although the lowest two moments of the Hamiltonian remain close to their initial values (not shown here). Higher moments of the Hamiltonian also exhibits notable deviation during the time evolution. Hence, for initial states that are far from a Gaussian distribution, we expect that controlling the lowest two moments may not be sufficient for ADA-Trotter to generate accurate digital simulation.

\begin{figure}
	\centering
	\includegraphics[width=0.9\linewidth]{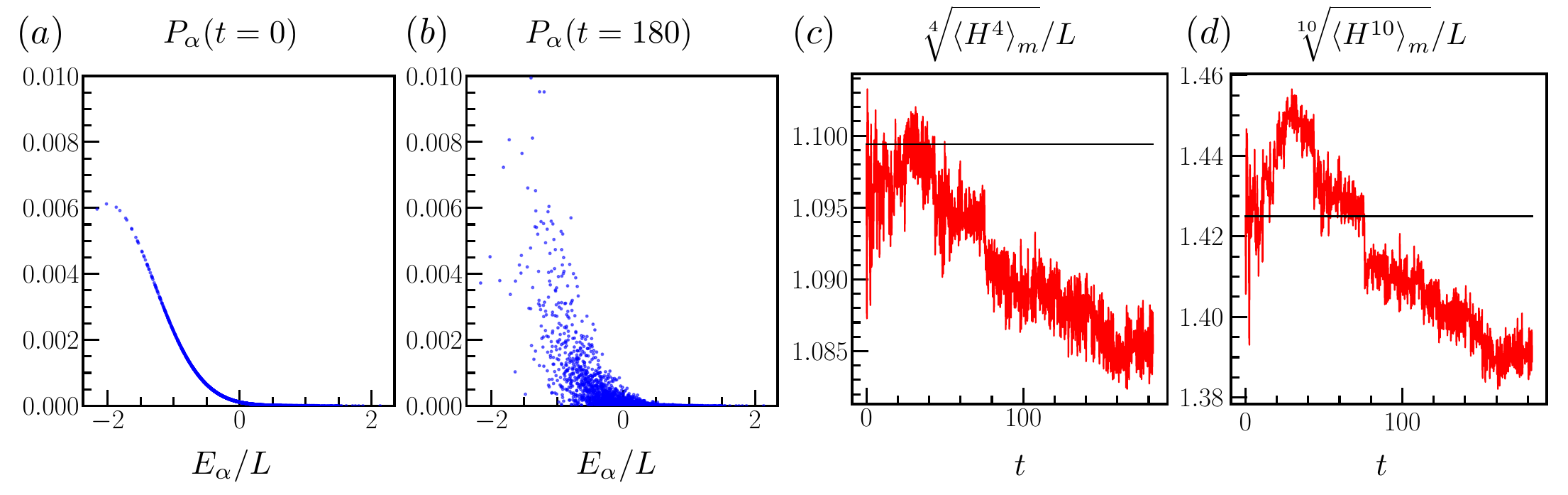}
	\caption{Energy distribution and higher moments of the Hamiltonian for an initial state that is restricted to low energy subspace. (a) and (b) show the energy distribution at early ($t=0$) and long times ($t=180$). (c) and (d) depict the higher moments where deviation from the exact conserved quantity (black) can be observed. We use $J_{z}=-1, h_x=-2,h_z=0.2, d_{\mathcal{E}}=0.01,d_{\delta\mathcal{E}^2}=0.1,L=16$ for numerical simulation.}
	\label{fig:powersH_lowT}
\end{figure}

Finally, we consider another initial state that is far from a Gaussian distribution, $\ket{\psi(0)} =(1/C)\sum_{\alpha}g(E_1,E_{\alpha})\ket{E_{\alpha}}$, with the normalization factor $C$ and $E_1=-2L, w=4\sqrt{2L}.$ As $E_1$ is sufficiently close to the ground state energy, $\ket{\psi(0)}$ is therefore restricted to the low energy subspace as shown in Fig.~\ref{fig:powersH_lowT} (a). After sufficiently long-time evolution, $P_{\alpha}$ deviates from the initial distribution, and higher moments also deviate from their initial values. However, compared with Fig.~\ref{fig:powersH_Cat}, errors in $P_{\alpha}$ and higher moments are now much smaller. The concentration at low energies may account for such behavior, as Trotterization induced perturbations may not be able to efficiently couple the low-energy eigenstates to other higher energies.

\subsection{Constrained heating with tight variance constraint}
\label{sec.long-time-dynamics} 
\begin{figure}
	\centering
	\includegraphics[width=0.6\linewidth]{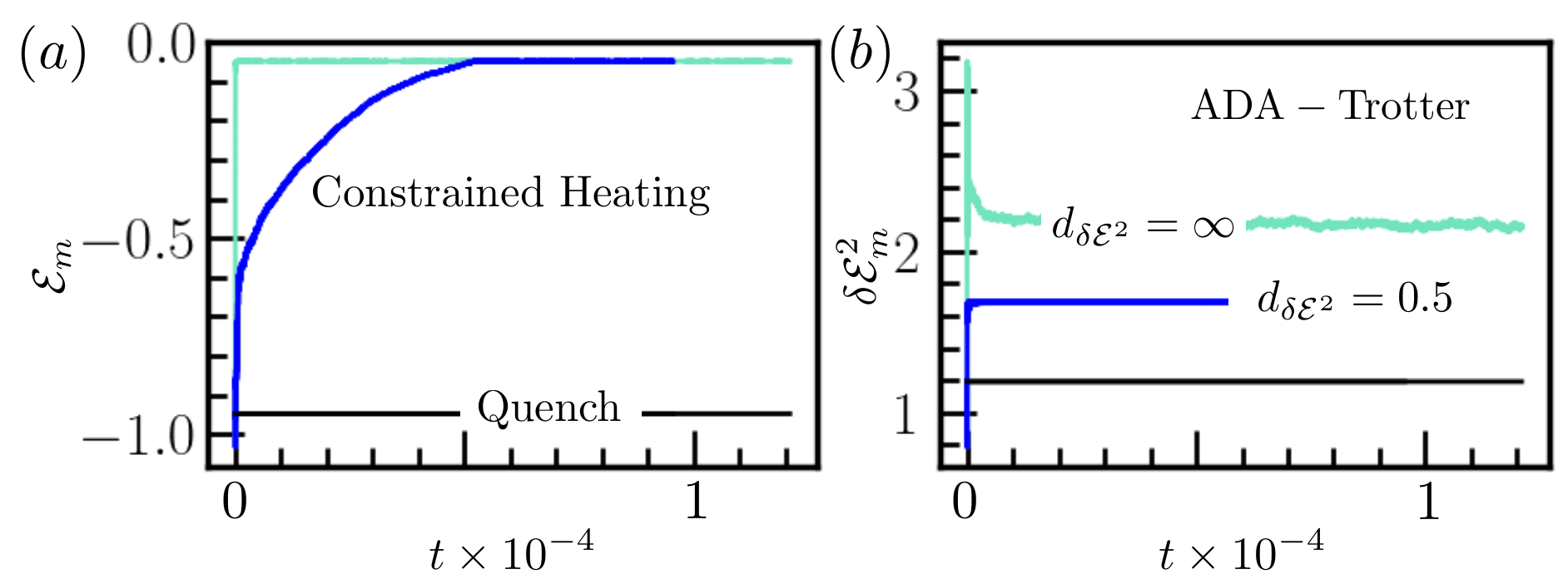}
	\caption{Dynamics of energy and its variance. Energy quickly heats up if variance is not constrained. The energy evolution slows down (blue) if energy variance is enforced to stay close to the initial value. We use $J_z=1,h_x=1,h_z=0.3, L=20$, and initial state $\exp(-i\pi/5\sum_j\sigma_j^y)\ket{\downarrow\dots \downarrow}$ for numerical simulation.}
	\label{fig:constrained_heating_SM}
\end{figure}
In the main text we assume that ADA-Trotter generates a diagonal ensemble with a shifted energy and energy variance at long times. Interestingly, to reach this long time limit, heating in energy can be highly constrained if the energy variance is enforced to stay close to its initial value. As shown in Fig.~\ref{fig:constrained_heating_SM}, we plot the time evolution in energy with different variance constraints. Without the variance constraint (green), the system heats up very quickly with the energy towards 0, corresponding to the infinite temperature. It does not eventually reach zero because here we also constrain the energy deviation with $d_{\mathcal{E}}=0.9.$ In contrast, with a tight variance constraint (blue), although the system eventually heats up to the same energy, the heating rate is significantly smaller by orders of magnitude than the green one. It is worthwhile to further explore the dependence of the constrained heating rate on the variance tolerance $d_{\delta \mathcal{E}^2}$, which may also provide useful insights on prethermalization in Floquet systems.

\subsection{Asymptotic error at long times for Fig.~\ref{fig:z_error}}
\label{sec:longtimeerror}
\begin{figure}[H]
	\centering
	\includegraphics[width=0.85\linewidth]{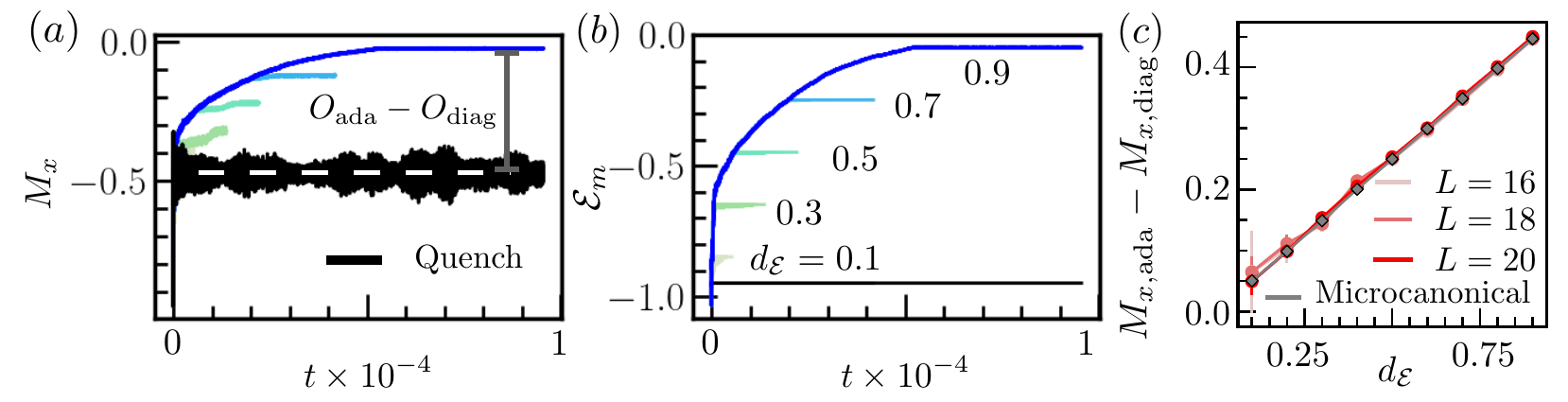}
	\caption{ Long time dynamics in local observable (a) and energy (b) for different energy constraints as labeled in panel (b). A smaller tolerance in energy leads to more accurate long time relaxation of local observables. (c) Error in local observables versus energy constraint. A linear dependence appears because the  micro-canonical value has a vanishing second derivative versus energy, $O''(\mathcal{E})=0$.
		We use $J_z=1,h_x=1,h_z=0.3$ and initial state $\exp(-i\pi/5\sum_j\sigma_j^y)\ket{\downarrow\dots \downarrow}$ for numerical simulation.
	}
	\label{fig:long_time_dynamics}
\end{figure}
Employing a smaller energy constraint can suppress the asymptotic error in local observables. In Fig.~\ref{fig:long_time_dynamics} (b), we use the same variance control $d_{\delta\mathcal{E}^2}$ but different energy constraints. The magnetization in $x$ direction is plotted in panel (a), where the error in local observable at long times increases with $d_{\mathcal{E}}$. The dependence is shown in panel (c) where the error is plotted for different system sizes. Unlike the error in z component in Fig.~\ref{fig:z_error} in the main text, here it scales linearly  with the tolerance. According to Eq.~\ref{eq.O_AE_d_thermo}, this observation suggests that the  micro-canonical prediction has a vanishing second derivatives versus energy, $O''(\mathcal{E})=0$. We also verify this by doing exact diagonalization of the target Hamiltonian and calculating the micro-canonical value for the local observable (grey dots in panel (c)).

\begin{figure}[h]
	\centering
	\includegraphics[width=0.55\linewidth]{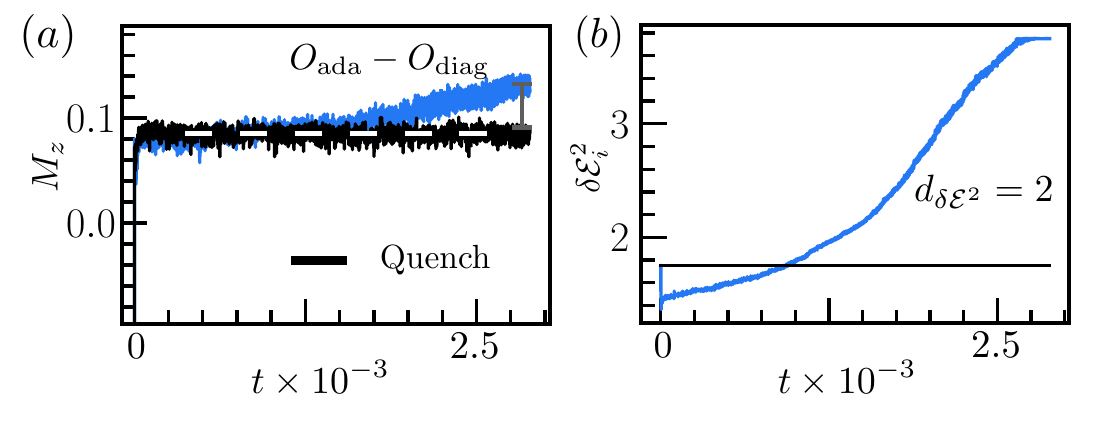}
	\caption{ Dynamics of local observable (a) and energy variance (b) with a tight energy constraint $d_{\mathcal{E}}=0.001$. Local observable exhibits a drift at later times, which is induced by the notable variance deviation. The final saturation of the variance deviation is bounded by $d_{\delta \mathcal{E}^2} = 2$. Such a drift is a finite size effect as discussed in the main text. We use $J_z=1,h_x=1,h_z=0.3$ and initial state $\exp(-i\pi/30\sum_j\sigma_j^y)\ket{\downarrow\dots \downarrow}$ for numerical simulation.
	}
	\label{fig:long_time_dynamics_variance}
\end{figure}
We then plot the long time evolution in the presence of a tight energy constraint ($d_{\mathcal{E}}=0.001$) in Fig.~\ref{fig:long_time_dynamics_variance}. There is a clear drift of the local observable in panel (a) at later times, induced by the notable deviation in the energy variance shown in panel (b). Such drift is a finite size effect as discussed in the main text and illustrated in Fig.~\ref{fig:z_error}(b).

\end{document}